\documentclass[11pt]{article}
\usepackage{amsfonts}
\usepackage{amsthm}
 \usepackage{bezier}

\usepackage{epic,eepic,amsmath,amssymb,color,graphicx}

\def\be{\begin{equation}}
\def\ee{\end{equation}}
\def\bea{\begin{eqnarray}}
\def\eea{\end{eqnarray}}

\newcommand{\sect}[1]{\setcounter{equation}{0}\section{#1}}
\newcommand{\subsect}[1]{\subsection{#1}}

\newcommand{\bq}{\mathbf{q}}
\newcommand{\bp}{\mathbf{p}}
 
 \newcommand{\kk}{\kappa}
 \newcommand{\del}{\delta}
  \newcommand{\Om}{\Omega}
 \newcommand{\la}{\lambda}
\newcommand{\te}{\phi}

 \newcommand{\non}{ {\cal H}^\prime}
\newcommand{\ii}{ {\cal I}^\prime}
\newcommand{\ele}{ {\cal L}^\prime}

\newcommand{\ji}{{q}}
\newcommand{\SW}{{\rm SW}}

\newcommand{\Sk}{{\rm\ \!S}}            
\newcommand{\Ck}{{\rm\ \!C}}           
\newcommand{\Tk}{{\rm\ \!T}}

\def\1{\'{\i}}                           

  \def\>#1{{\mathbf#1}}                 
\def\RR{\mathbb{R}}

 \def\m{\mu}
\def\k{{\kappa}}

\begin{document}

\thispagestyle{empty}

  \noindent
 {\Large{\bf {The anisotropic oscillator on the  two-dimensional sphere \\[6pt]    and the  hyperbolic plane}}}

\medskip 
\medskip 
\medskip

\begin{center}
{\sc \'Angel Ballesteros$^1$, Francisco J. Herranz$^1$ and Fabio Musso$^{2}$}
\end{center}

\noindent
{$^1$ Departamento de F\'\i sica,  Universidad de Burgos,
E-09001 Burgos, Spain  \\[10pt] 
$^2$ Dipartimento di Fisica  `Edoardo Amaldi',  Universit\'a Roma Tre,
I-00146 Rome, Italy \\
}

\noindent
 E-mail: {\tt   angelb@ubu.es,   fjherranz@ubu.es, fmusso@ubu.es}

  \medskip 
\bigskip
\bigskip

\begin{abstract} 
\noindent
An integrable generalization on the two-dimensional sphere ${\mathbf S}^2$ and the hyperbolic plane ${\mathbf H}^2$ of the Euclidean anisotropic oscillator Hamiltonian with `centrifugal'  terms given by
\begin{equation}
\mathcal{H}=\frac{1}{2}(p_1^2+p_2^2)+ \delta q_{1}^{2}+(\delta + \Omega)q_{2}^{2} +\frac{\lambda_1}{q_{1}^{2}}+\frac{\lambda_2}{q_{2}^{2}}
\nonumber
\end{equation}
 is presented. The resulting   generalized Hamiltonian ${\cal H}_\kk$   depends explicitly on the constant Gaussian curvature $\kk$ of the underlying space, in such a way that   all the results here presented    hold simultaneously for ${\mathbf S}^2$  $(\k>0)$,  ${\mathbf H}^2$  $(\k<0)$ and  ${\mathbf E}^2$  $(\k=0)$. Moreover, ${\cal H}_\kk$  is explicitly shown to be integrable for any values of the parameters $\delta$, $\Omega$, $ \lambda_1$ and $\lambda_2$. Therefore, ${\cal H}_\kk$ can also be  interpreted as an anisotropic generalization of the curved Higgs oscillator, that is recovered as the isotropic limit $\Om\to0$ of ${\cal H}_\kk$. Furthermore,  numerical integration of some of the trajectories for ${\cal H}_\kk$ are worked out and the dynamical features arising from the introduction of a curved background are highlighted. 
 
 The superintegrability issue for ${\cal H}_\kk$ is discussed by focusing on the value $\Om=3\del$,   which is one of the cases for which the Euclidean Hamiltonian $\mathcal{H}$ is known to be superintegrable (the 1:2  oscillator). We show numerically that for $\Om=3\delta$ the curved Hamiltonian ${\cal H}_\kk$  presents nonperiodic bounded trajectories, which seems to indicate that ${\cal H}_\kk$ provides a non-superintegrable generalization of ${\cal H}$ even for values of $\Om$ that lead to commensurate frequencies in the Euclidean case. We compare this result with a previously known superintegrable curved analogue ${\cal H}_\kk'$ of the 1:2 Euclidean oscillator, which is described in detail, 
 showing that the $\Om=3\del$ specialization of ${\cal H}_\kk$ does not coincide with ${\cal H}_\kk'$. Hence we conjecture that ${\cal H}_\kk$ would be an integrable (but not superintegrable) curved generalization of the anisotropic oscillator that exists for any value of $\Om$ and has constants of the motion that are quadratic in the momenta. Thus each commensurate Euclidean oscillator could admit another specific superintegrable curved Hamiltonian which would be different from ${\cal H}_\kk$ and endowed with higher order integrals. Finally, the geometrical interpretation of the curved  `centrifugal' terms appearing in ${\cal H}_\kk$ is also   discussed in detail.

\end{abstract}

\bigskip\bigskip\bigskip\bigskip


\bigskip


\newpage


\sect{Introduction}

The two-dimensional (2D) anisotropic oscillator with `centrifugal' (or  `Rosochatius') terms
\be
\mathcal{H}=\frac{1}{2}(p_1^2+p_2^2)+ \del q_{1}^{2}+(\del + \Om)q_{2}^{2} +\frac{\la_1}{q_{1}^{2}}+\frac{\la_2}{q_{2}^{2}}
\label{ba},
\ee
with $\del$, $\Om$ and $\la_i$ being real parameters and $\{q_i,p_j\}=\delta_{ij}$ $(i,j=1,2)$,
is one of  the most elementary (albeit instructive) examples of completely integrable Hamiltonian systems. 
Since integrability will be our essential guiding principle when constructing the curved generalization of this system, let us firstly recall the (Liouville) integrability properties of $\mathcal{H}$ in terms of the $\del$, $\Om$ and $\la_i$ parameters. 

Obviously,  $\mathcal{H}$ is  always integrable for any value of $\Om$ since it is separable in Cartesian coordinates:
\be
{\cal H}={\cal I}_1+{\cal I}_2,\qquad 
 {\cal I}_1=\frac{1}{2} p_1^2 +\del  q_{1}^{2}  +\frac{\la_1}{q_{1}^{2}},
\qquad {\cal I}_2=\frac{1}{2} p_2^2 +(\del +\Om) q_{2}^{2} +\frac{\la_2}{q_{2}^{2}}.
\label{bb}
\ee
Thus each pair  $({\cal H},{\cal I}_1)$ and $({\cal H},{\cal I}_2)$ is formed by two functionally independent functions in involution with respect to the canonical Poisson bracket. However, there are two particular situations for which (\ref{ba}) is known to be superintegrable due to the existence of an additional independent integral of the motion:

\begin{itemize}
\item  When $\Om=0$.  This is  is just the isotropic oscillator with Rosochatius terms ($\la_i$ arbitrary), which is frequently called the anisotropic singular oscillator or the Smorodinsky--Winternitz (SW) system~\cite{FMSUW65, Ev90b, 8, 10}. In this case the additional integral of the motion is quadratic in the momenta and reads:
$$
{\cal J}=({q_1}{p_2} -
{q_2}{p_1})^2 +2 \left(
\la_1\frac{q_2^2}{q_1^2}+\la_2\frac{q_1^2}{q_2^2}\right).
$$

\item  If the value of $\Om$ is such that  the ratio of the associated oscillator frequencies is a rational number ($\la_i$ can   be  arbitrary)~\cite{Jauch, Tempesta}. In this situation the only case in which the system presents an additional integral that is {\em quadratic} in the momenta is the commensurate 1:2 oscillator ($\Om=3\del$) plus a  {\em single} Rosochatius potential (say $\la_2=0$). If  either both $\la_i\ne 0$ or  the anisotropic   oscillator   has  any other pair of commensurate frequencies then the additional integral is of higher order in the momenta~\cite{Jauch, Tempesta}. The resulting 1:2 superintegrable Hamiltonian $ \non$ and its  quadratic integrals of motion turn out to be  
\bea
&&  \non=\frac{1}{2} (p_1^2+p_2^2)+ \del q_{1}^{2}+4\del q_{2}^{2} +\frac{\la_1}{q_{1}^{2}} 
 ,\qquad \nonumber\\[2pt]
&& \ii_1\equiv {\cal I}_1= \frac{1}{2} p_1^2+ \del q_{1}^{2}  +\frac{\la_1}{q_{1}^{2}} ,
\qquad \ii_2=\frac{1}{2} p_2^2 +4 \del q_{2}^{2}  ,\label{bd}
\\
&& \ele= (q_1p_2-q_2p_1)p_1+    2 \del  q_{1}^{2} q_2 - \frac{2 \la_1 q_2}{q_{1}^{2}}.
\nonumber
\eea
Again, $\non= \ii_1+\ii_2$ and it can be straightforwardly checked that the sets  $(\non,\ii_1,\ele)$ and $(\non,\ii_2,\ele)$ are formed by three functionally independent functions. 
\end{itemize}

The aim of this paper is to present a   generalization ${\cal H}_\kk$ of the Hamiltonian (\ref{ba}) on the 2D spaces with  constant Gaussian  curvature $\kk$.
Such a new Hamiltonian ${\cal H}_\kk$ can properly be  considered as a `curved' generalization of $\mathcal{H}$ since it presents two outstanding features:

\begin{itemize}
\item   ${\cal H}_\kk$ depends explicitly on the parameters $\kk$, $\del$, $\Om$ and $\la_i$, and is shown to be integrable for any value of them.

\item The Euclidean limit $\kk\to 0$ of ${\cal H}_\kk$  is well defined and leads to the Hamiltonian $\mathcal{H}$. 

\end{itemize}

An essential point of our approach consists in making use of the curvature $\k$  of the underlying 2D space as an explicit deformation parameter. Therefore,  
 all the results here presented will  be simultaneously valid for the 2D sphere ${\mathbf S}^2$ $(\k>0)$, the hyperbolic plane  ${\mathbf H}^2$  $(\k<0)$ and  the Euclidean plane ${\mathbf E}^2$  $(\k=0)$, and  the 
 `flat'  limit $\k\to 0$ in the spherical and hyperbolic systems will  lead to the corresponding Euclidean Hamiltonian~\eqref{ba}. Conversely, the non-trivial dynamical nature of the transition from the `flat'  systems defined on   ${\mathbf E}^2$ to the  `curved' ones associated to 
 ${\mathbf S}^2$ and  ${\mathbf H}^2$ will be  clearly    appreciated through this procedure, since ${\cal H}_\kk$ can be thought of as an integrable deformation  of ${\cal H}$. 

The novelty of ${\cal H}_\kk$ stems from the fact that, to the best of our knowledge, there are only two specific values of $\Om$ in  (\ref{ba}) for which  an integrable curved generalization of the corresponding Euclidean oscillator is known in the literature. Moreover, both  known curved systems are superintegrable with an additional integral that is quadratic in the momenta, and correspond with the two Euclidean cases that have been explicitly described above. These two known curved Hamiltonians are:

\begin{itemize}
\item  The so-called Higgs oscillator~\cite{Higgs,Leemon}, which is the curved analogue of the isotropic ($\Om=0$) Euclidean oscillator. In this case a curved superintegrable generalization of both  centrifugal terms can be added to the oscillator potential, and the resulting Hamiltonian is called the SW system on the 2D sphere and the hyperbolic plane~\cite{Pogosyan1,Pogosyan2,RS,int,kiev}.

\item A curved analogue of the nonisotropic 1:2  system (the $\Om=3\delta$ Euclidean oscillator) but with only one centrifugal term ($\la_2=0$), which was found in~\cite{RS}. We will call this system ${\cal H}_\kk'$. Note that if   we want  a second curved centrifugal term ($\la_2\neq 0$)     to be included without loosing   superintegrability, the additional curved integral should be of higher order in the momenta. This generalization is not known.

\end{itemize}

Therefore, a  `generic'  ({\em i.e.}, with $\Omega$ as an explicit parameter) integrable generalization ${\cal H}_\kk$ of (\ref{ba}) on ${\mathbf S}^2$ and  ${\mathbf H}^2$ was lacking, and we present it in this paper. Furthermore, the integrals of the motion for ${\cal H}_\kk$ will be quadratic in the momenta, and since the limit $\Omega\to 0$ of ${\cal H}_\kk$ will  lead to the Higgs oscillator, we  shall call ${\cal H}_\kk$ as the  `anisotropic Higgs oscillator'. On the contrary, the known superintegrable system ${\cal H}_\kk'$ is by no means the particular $\Om=3\delta$ case in the $\Om$-dependent family  ${\cal H}_\kk$, and has to be considered as an `isolated' system that deserves a separate analysis.

In this context, the natural question concerning the superintegrability of ${\cal H}_\kk$ arises. As we have mentioned before, ${\cal H}_\kk$ can be considered as an integrable deformation  of ${\cal H}$ in terms of the curvature parameter $\kk$. However, the Euclidean system ${\cal H}$ is not superintegrable for a generic value of $\Om$, since  it is well known that all the bounded trajectories for ${\cal H}$ are periodic (Lissajous curves) only in the commensurate and $\Omega=0$ cases. Therefore, one should not expect ${\cal H}_\kk$ to be superintegrable for an arbitary value of $\Om$.
In this respect, we have performed a numerical analysis of bounded trajectories of the Hamiltonian ${\cal H}_\kk$ for   $\Om=0$ as well as for different values of $\Om$ leading to commensurate frequencies in the Euclidean system. In all the cases we have also considered several different initial conditions, and the result is that ${\cal H}_\kk$ seems to be  superintegrable {\em only in the $\Omega=0$ case} (which is the isotropic Higgs oscillator or, equivalently, the 1:1 oscillator) since only in this case bounded trajectories for ${\cal H}_\kk$ are always found to be periodic.
In particular, we illustrate this analysis by presenting some bounded trajectories for ${\cal H}_\kk$ with $\Om=3\del$ and by comparing them with the ones of ${\cal H}_\kk'$, which are found to be periodic in agreement with the well known superintegrability of the latter (note that the neat periodicity of these trajectories gives suppport to the stability of the numerical analysis performed in all the remaining cases).

All these results  suggest that ${\cal H}_\kk$ would be the generic $\Om$-dependent integrable (but not superintegrable) curved generalization of the anisotropic oscillator, and that each commensurate Euclidean oscillator could admit another curved generalization that would be given in each case by a different Hamiltonian. This conjecture would mean, for instance, that the curved analogue of the 1:3 oscillator ($\Om=8\delta$) would be a new Hamiltonian ${\cal H}_\kk''$ which would be indeed different from the system obtained from ${\cal H}_\kk$ by taking $\Om=8\delta$ (note that the additional integral for ${\cal H}_\kk''$ has to be of higher order in the momenta). Nevertheless, both Hamiltonians would  `collapse'  to a common system in the Euclidean limit $\kk\to 0$.

The paper is organized as follows. In the next section we briefly recall the basics on Beltrami and  geodesic polar variables that generalize  the Cartesian and polar ones, respectively, to  ${\mathbf S}^2$  and  ${\mathbf H}^2$. In section 3 we introduce the anisotropic Higgs oscillator ${\cal H}_\kk$ and we give the explicit form of its integral of the motion for generic values of all the parameters, including $\Om$. We also present some numerical analysis of its trajectories for some commensurate values of its Euclidean frequencies, showing that superintegrability does not seem to survive under the deformation introduced through the curvature. In section 4 we analyse the superintegrable curved oscillator given by ${\cal H}_\kk'$, and the 1:2 Lissajous curves on ${\mathbf S}^2$  and  ${\mathbf H}^2$ are found as the trajectories for the system.
In section 5 we perform a detailed geometrical  analysis of the  `curved' centrifugal terms~\cite{ann0} of all the previous curved oscillators, and their dynamical effects  are also numerically explored.
Finally, the last section is devoted to some concluding remarks and open problems.


\sect{Geodesic polar and  Beltrami    variables}

First of all, let us rewrite the Euclidean Hamiltonians given in the previous section in terms of polar coordinates $(r,\te)$ with conjugate momenta $(p_r,p_\te)$. This change of coordinates induces the following canonical transformation 
\be
 q_1= r \cos\te,\quad q_2=r\sin\te ,\quad p_1= \cos\te\,p_r-\frac{\sin\te}{r}\, p_\te,\quad p_2= 
 \sin\te\,p_r+\frac{\cos\te}{r}\, p_\te ,
 \label{be}
\ee
that transforms the expressions (\ref{ba})--(\ref{bd}) into
\bea
&& {\cal H}=\frac{1}{2}\left(p_{r}^{2}+\frac{p_{\te}^{2}}{r^2} \right)+ \del r^{2}+\Om\, r^2 \sin^2\te  +\frac{\la_1}{r^{2}\cos^2\te}  +\frac{\la_2}{r^{2}\sin^2\te} 
,\nonumber\\[2pt]
&&{\cal I}_1= \frac 12 \left( \cos\te\,p_r-\frac{\sin\te}{r}\, p_\te\right)^2 +\del r^2 \cos^2\te +\frac{\la_1}{r^2\cos^2\te}, \label{bf}\\[2pt]
&&{\cal I}_2= \frac 12 \left(  \sin\te\,p_r+\frac{\cos\te}{r}\, p_\te \right)^2 +(\del+\Om)  r^2 \sin^2\te +\frac{\la_2}{r^2\sin^2\te} ;
\nonumber
\eea
 \bea
&& \non=\frac{1}{2}\left(p_{r}^{2}+\frac{p_{\te}^{2}}{r^2} \right)+ \del r^{2}+3\del\, r^2 \sin^2\te  +\frac{\la_1}{r^{2}\cos^2\te}   ,\nonumber\\[2pt]
&& \ii_1\equiv {\cal I}_1,\qquad \ii_2= \frac 12 \left(  \sin\te\,p_r+\frac{\cos\te}{r}\, p_\te \right)^2 +4\del  r^2 \sin^2\te,
\label{bg}\\[2pt]
&&\ele=  \left( \cos\te\,p_r-\frac{\sin\te}{r}\, p_\te\right)  p_\te +2 \del r^3 \cos^2\te\sin\te -\frac{2\la_1\tan\te}{r\cos\te}  .
\nonumber
\eea

Now, in order to generalize the  two  above  {\em flat}   systems (\ref{bf}) and (\ref{bg})  to the   sphere and  to the hyperbolic plane    we will apply a geometric aproach in which ${\mathbf S}^2$ and ${\mathbf H}^2$ are obtained as homogeneous spaces of certain 3D Lie groups. In this way, the realization of the associated symmetry generators in terms of  geodesic polar and Beltrami (projective)  coordinates will enable us to  propose the curved counterparts of these two systems.

\subsection{Geodesic polar coordinates}

Let $\mathfrak{so}_\kk(3)$  be the 3D real Lie algebra with generators   $J_{01}$, $J_{02}$ and $J_{12}$ 
  with commutation relations and Casimir operator given by
 \be
  [J_{12},J_{01}]=J_{02},\quad [J_{12},J_{02}]=-J_{01},\quad [J_{01},J_{02}]=\kk J_{12} ,\quad
  {\cal C}=J_{01}^2+J_{02}^2+\kk J_{12}^2,
 \label{ca}
 \ee
  where $\kk$ is a real parameter.
The three 2D classical Riemannian spaces of constant Gaussian curvature $\kk$ can  be collectively constructed as the homogenous spaces $SO_\kk(3)/SO(2)$  where   $SO_\kk(3)$ is the Lie group generated by $\mathfrak{so}_\kk(3)$ and 
$SO(2)$ is the isotopy subgroup spanned by $J_{12}$. Hence $J_{01}$ and $J_{02}$ play the role of    generators of translations, while $J_{12}$ corresponds to the generator of rotations on the 2D space. In this way, according to the value of $\kk$ we get   the  three 
 particular homogeneous spaces:
  $$
\begin{array}{lll}
\kk>0:\ \mbox{Sphere}&\qquad \kk=0:\ \mbox{Euclidean plane}&\qquad \kk<0:\ \mbox{Hyperbolic plane}\\[2pt]
 {\mathbf S}^2=SO(3)/SO(2)&\qquad  {\mathbf E}^2=ISO(2)/SO(2)&\qquad   {\mathbf H}^2=SO(2,1)/SO(2)
\end{array}
\label{cb}
$$
 We recall that $J_{12}$ leaves a point $O$ invariant,   the origin, and that each   translation generator $J_{0i}$ moves   $O$  along a base geodesic  $l_i$ in such a manner that $l_1$ and $l_2$ are orthogonal at $O$ as depicted in figure 1.

These   spaces can    be globally  embedded  in the  linear space $\mathbb R^{3}$ with
  ambient  or Weierstrass coordinates $(x_0,\>x)=(x_0,x_1,x_2)$ by imposing  the  `sphere'
constraint $\Sigma$: $x_0^2+ \kk\,  \>x^2=1$; the origin corresponds to   the point $O=(1,\>0)\in \mathbb R^{3}$.
The geodesic polar coordinates $r\in (0,\infty)$ and $\te\in[0,2\pi)$  are defined as follows.  If the particle is located at a point $Q$, 
then $r$ is   the distance between  $Q$ and the origin $O$ measured along the  
geodesic $l$ that joins both points. On the other hand, $\te$ is  the angle which  determines the orientation of $l$ with respect to the base geodesic $l_1$  (see  figure 1).
The ambient coordinates  are parametrized in terms of $(r,\te)$ in the form~\cite{RS,conf}
$$
x_0=\Ck_\kk(r),\qquad x_1=\Sk_\kk(r)\cos\te , \qquad x_2=\Sk_\kk(r)\sin\te,
\label{cf}
$$
where we have introduced the curvature-dependent functions
$$
\Ck_{\kk}(r) 
=\left\{
\begin{array}{ll}
  \cos {\sqrt{\kk}\, r} &\   \kk >0 \cr 
  1  &\  \kk  =0 \cr 
\cosh {\sqrt{-\kk}\, r} &\  \kk <0 
\end{array}\right. ,\qquad 
\Sk_{\kk}(r) 
=\left\{
\begin{array}{ll}
    \frac{1}{\sqrt{\kk}} \sin {\sqrt{\kk}\, r} &\   \kk >0 \cr 
  r &\  \kk  =0 \cr 
\frac{1}{\sqrt{-\kk}} \sinh {\sqrt{-\kk}\, r} &\  \kk <0 
\end{array}\right.  .
\label{cg}
$$
The $\kk$-tangent is defined by $\Tk_{\kk}(r) = {\Sk_\kk(r)}/{ \Ck_\kk(r)}$.
Properties  for these $\kk$-functions can be found in~\cite{RS,conf,trigo}. For instance,
$$
\Ck^2_\kk(r)+\kk\Sk^2_\kk(r)=1,\qquad \frac{ {\rm d}}
{{\rm d} r}\Ck_\kk(r)=-\kk\Sk_\kk(r),\qquad 
\frac{ {\rm d}}
{{\rm d} r}\Sk_\k(r)= \Ck_\kk(r) .
$$

 Now let us denote by $(p_r,p_\te)$ the conjugate momenta of $(r,\te)$. The canonical Poisson bracket for the three functions 
 \be
J_{01}=  \cos\te\,p_r-\frac{\sin\te}{ \Tk_\kk(r) }\, p_\te,\qquad 
J_{02}= \sin\te\,p_r+\frac{\cos\te}{\Tk_\kk(r)}\, p_\te ,\qquad J_{12}=p_\te 
    \label{ch}
  \ee
gives us the Lie--Poisson analogue of the algebra  $\mathfrak{so}_\kk(3)$  (\ref{ca}). Therefore, (\ref{ch}) provides a symplectic realization of the Lie--Poisson analogue of $\mathfrak{so}_\kk(3)$, 
and we can write the kinetic energy ${\cal T}$ for the free motion of a particle moving on  ${\mathbf S}^2$ and ${\mathbf H}^2$ as the above symplectic realization for the Casimir function of $\mathfrak{so}_\kk(3)$, namely:
\be
 {\cal T}\equiv\frac 12 {\cal C}=\frac 12 (J_{01}^2+J_{02}^2+\kk J_{12}^2)=\frac 12 \left( p_r^2+\frac{p_\te^2}{\Sk^2_\kk(r)} \right) .
 \label{ci}
 \ee
Obviously, the $\kk\to 0$ limit of this expression gives the 2D Euclidean kinetic energy.


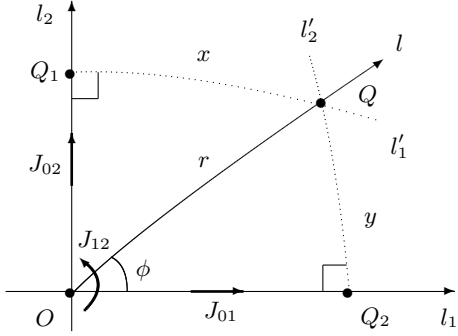
\begin{figure}[t]
\begin{center}
\begin{picture}(170,145)
\footnotesize{
\put(127,22){$\bullet$}
\put(52,33){\makebox(0,0){$\te$}}
 \qbezier(46,25)(46,35)(40,38)
\put(138,52){\makebox(0,0){$y$}}
\put(75,114){\makebox(0,0){$x$}}
\put(75,74){\makebox(0,0){$r$}}
\put(142,112){\vector(4,3){1}}
\put(117,94){$\bullet$}
\put(140,15){\makebox(0,0){$Q_2$}}
\put(137,99){\makebox(0,0){$Q$}}
\put(149,80){\makebox(0,0){$l'_1$}}
\put(149,119){\makebox(0,0){$l$}}
\put(22,22){$\bullet$}
\put(15,15){\makebox(0,0){$O$}}
\put(22,105){$\bullet$}
\put(15,108){\makebox(0,0){$Q_1$}}
\put(15,130){\makebox(0,0){$l_2$}}
\put(25,10){\vector(0,1){125}}
\put(0,25){\vector(1,0){170}}
\put(168,15){\makebox(0,0){$l_1$}}
\put(35,98){\line(0,1){10}}
\put(25,98){\line(1,0){10}}
\put(120,25){\line(0,1){10}}
\put(120,35){\line(1,0){9}}
\qbezier[50](25,108)(70,110)(140,90)
\qbezier[50](130,25)(125,80)(115,114)
\put(115,124){\makebox(0,0){$l'_2$}}
\qbezier(25,24)(50,50)(140,111)
\linethickness{1pt}
\put(70,25){\vector(1,0){20}}
\put(25,65){\vector(0,1){20}}
\put(15,73){\makebox(0,0){$J_{02}$}}
\qbezier(30,18)(40,28)(30,36)
\put(29,37){\vector(-1,1){1}}
\put(75,15){$J_{01}$}
\put(27,42){$J_{12}$}
}
\end{picture}
\end{center}
\noindent
\\[-40pt]
\caption{\footnotesize Schematic representation of the symmetry generators $J_{ij}$ and geodesic distances  of a   point $Q$ with geodesic polar coordinates $(r,\te)$ on a  2D   homogeneous space.}
\label{figure4}
\end{figure}


\subsection{Beltrami coordinates}

On the other hand,  if  we apply the central projection with pole  
$(0,\>0)\in \mathbb R^{3}$  from  $ (x_0,\>x)$
to the Beltrami coordinates  $\>\ji\in \mathbb R^2$,  that are defined as
$(0,\>0)+\m\,
(1,\>q)\in\Sigma$, 
we get the relations
$$
\m=\frac{1}{\sqrt{1+ \kk \>q^2}},\qquad
 x_0=\m ,\qquad 
\>x=\m\, \>q=\frac{\>q}{\sqrt{1+ \k \>q^2}}.
\label{cc}
$$
The image of this  projection     is the subset of   $\mathbb  R^3$   with    $\mu\in \RR$ so that   $1+ \k \>q^2>0$. In the hyperbolic or Lobachevsky plane with $\k=-|\k|<0$ this gives the interior of the Poincar\'e disk: $1>|\k|  \>q^2$.
Notice that when   $\k=0$, the Beltrami coordinates reduce to the Cartesian ones on $\mathbf E^2$: $x_0=1$ and
   $\>x=  \>q$. Hereafter we assume the following notation for    $\bq=(q_1,q_2)$ and for the conjugate momenta  $\bp=(p_1,p_2)$:  
$$
\bq^2=q_1^2+q_2^2,\qquad \bp^2=p_1^2+p_2^2,\qquad \bq\cdot \bp=q_1 p_1+ q_2 p_2,\qquad  |\bq|=\sqrt{q_1^2+q_2^2}  .
\label{bba}
$$
In terms of this second set of canonical variables,
another symplectic realization of the Lie--Poisson algebra
 $\mathfrak{so}_\kk(3)$ is found to be~\cite{Kepler}
 \be
J_{0i}= p_i+\kk (\bq\cdot\bp) q_i,   \quad i=1,2; \qquad J_{12}=q_1 p_2 - q_2 p_1 ,
  \label{cd}
 \ee
and the free Hamiltonian $ {\cal T}$  that determines the geodesic motion on the Riemannian  spaces is again obtained from the Casimir (\ref{ca}) under the above realization, namely
 \be
 {\cal T}\equiv\frac 12 {\cal C}=\frac 12 (J_{01}^2+J_{02}^2+\kk J_{12}^2)=\frac 12 \left(1+\kk \bq^2\right) \left(\bp^2+\kk(\bq\cdot\bp)^2 \right) .
 \label{ce}
 \ee

We remark that, as expected,  the Beltrami and geodesic polar   variables
are connected through   a canonical transformation which is given by
\be
\begin{array}{ll}
\displaystyle{  q_1=\Tk_\kk(r)\cos\te},&\quad \displaystyle{  p_1=\Ck^2_\kk(r)\cos\te \, p_r-\frac{\sin\te}{\Tk_\kk(r)}\,p_\te},\\[10pt]
\displaystyle{  q_2=\Tk_\kk(r)\sin\te},&\quad \displaystyle{  p_2=\Ck^2_\kk(r)\sin\te \,p_r+\frac{\cos\te}{\Tk_\kk(r)}\,p_\te},
  \end{array}
\label{cj}
\ee
and generalizes (\ref{be}). The inverse change of coordinates  leads to the expressions
\be
\begin{array}{lll}
\displaystyle{\Ck_\kk(r) =\frac{1}{\sqrt{1+\kk \bq^2}  }  },&\quad \displaystyle{   \Sk_\kk(r) =\frac{|\bq|}{\sqrt{1+\kk \bq^2}  }  },&\quad \displaystyle{   \Tk_\kk(r) = |\bq| } ,\\[10pt]
\displaystyle{  \cos\te=\frac{q_1}{|\bq| }    } ,&\quad \displaystyle{   \sin\te=\frac{q_2}{|\bq| }  },&\quad \displaystyle{    \tan\te=\frac{q_2}{q_1}} .\\[10pt]
\end{array}
\label{ck}
\ee


\sect{The anisotropic Higgs oscillator}

In order to generalize   the integrable anisotropic  Hamiltonian on ${\mathbf E}^2$ (\ref{bf})  to the sphere and the hyperbolic plane we recall that, in terms of geodesic polar coordinates, the Higgs oscillator potential on both spaces  is given by $\Tk^2_\kk(r)$~\cite{RS,int,kiev,Ranran} (i.e.~$\tan^2 r$ on ${\mathbf S}^2$ with $\k=1$ and $\tanh^2 r$ on  ${\mathbf H}^2$ with $\k=-1$), meanwhile  the   curved Kepler potential  is $1/\Tk_\kk(r)$ (a detailed study of the latter can be found in~\cite{Voz}).
Furthermore, the  superposition of the  Higgs 
oscillator with two   centrifugal potentials ({i.e.}~the curved SW system)
is also known in these variables  and defines   a superintegrable Hamiltonian.  Such a system will be our starting point, whose properties are summarized as follows.

\medskip

\noindent
{\bf Proposition 1.}~\cite{RS,int,kiev} {\em  The Higgs oscillator Hamiltonian on ${\mathbf S}^2$ and  ${\mathbf H}^2$ with two Rosochatius
terms (curved SW system) is given in terms of geodesic polar variables by
\be
{\cal H}_\SW=\frac 12 \left( p_r^2+\frac{p_\te^2}{\Sk^2_\kk(r)} \right) +\del \Tk^2_\kk(r) +\frac{\la_1}{\Sk^2_\kk(r) \cos^2\te}+\frac{\la_2}{\Sk^2_\kk(r) \sin^2\te} .
\label{dda}
\ee
The Hamiltonian ${\cal H}_\SW$ Poisson-commutes with the functions
 \bea
 &&  {I}_{01}=\frac{1}{2} J_{01}^2+\del  \Tk^2_\kk(r) \cos^2\te +\frac{\la_1}{\Tk^2_\kk(r) \cos^2\te},\nonumber\\[2pt]
 &&   {I}_{02}=\frac{1}{2}  J_{02}^2 +\del  \Tk^2_\kk(r) \sin^2\te   +\frac{\la_2}{\Tk^2_\kk(r) \sin^2\te} ,  \label{ddb}\\ 
   &&   {I}_{12}=\frac{1}{2}   J_{12}^2  + {\la_1}  \tan^2\te +  \frac {\la_2}{  \tan^2\te } ,
\nonumber
 \eea
where $J_{ij}$ are the functions  (\ref{ch}). Each set  $({\cal H}_\SW,   {I}_{01} ,  {I}_{12} )$ and $({\cal H}_\SW,   {I}_{02} ,  {I}_{12} )$ is formed by three functionally independent functions. Moreover, ${\cal H}_\SW= {I}_{01}+ {I}_{02}+ \k{I}_{12}+\k(\la_1+\la_2)$.
 }
\medskip

Essentially,    our task now consists in the introduction of an anharmonic $\Om$-term in the Hamiltonian (\ref{dda}) in such a way that the integrability   of the system is preserved. Obviously, 
this process requires to modify accordingly  the integrals of motion (\ref{ddb}). 
 The final result is presented in the following statement, that can be proven through direct computation.

\medskip

\noindent
{\bf Proposition 2.} {\em  The Hamiltonian
\be
{\cal H}_\kk=\frac 12 \left( p_r^2+\frac{p_\te^2}{\Sk^2_\kk(r)} \right) +\del \Tk^2_\kk(r)+\Om \Sk^2_\kk(r)\sin^2\te+\frac{\la_1}{\Sk^2_\kk(r) \cos^2\te}+\frac{\la_2}{\Sk^2_\kk(r) \sin^2\te} 
\label{da}
\ee
Poisson-commutes with the functions
 \bea
 &&  {\cal I}_{1,\kk}=\frac{1}{2} J_{01}^2+\del  \Tk^2_\kk(r) \cos^2\te +\frac{\la_1}{\Tk^2_\kk(r) \cos^2\te},\nonumber\\[2pt]
 &&   {\cal I}_{2,\kk}=\frac{1}{2} \left( J_{02}^2+\kk J_{12}^2 \right) +\del  \Tk^2_\kk(r) \sin^2\te  + \Om  \Sk^2_\kk(r) \sin^2\te +\frac{\la_2}{\Tk^2_\kk(r) \sin^2\te} \nonumber\\[2pt]
  &&\qquad\qquad\qquad + {\kk}   \left(  \frac  {\la_1 }{ \cos^2\te } + \frac  {\la_2 }{ \sin^2\te }    \right) ,  
  \label{db}
 \eea
where $J_{ij}$ are the functions  (\ref{ch}). The pairs  $({\cal H}_\kk,    {\cal I}_{1,\kk} )$ and $({\cal H}_\kk,    {\cal I}_{2,\kk} )$ are constituted by two functionally independent functions, and ${\cal H}_\kk={\cal I}_{1,\kk}+{\cal I}_{2,\kk}$. }
\medskip

Therefore, ${\cal H}_\kk$  (\ref{da}) can be considered as the anisotropic  Higgs oscillator Hamiltonian on ${\mathbf S}^2$ and  ${\mathbf H}^2$  with two Rosochatius
terms. We stress that for a generic value of $\Om$,  ${\cal H}_\kk$  is an integrable (but, presumably, non-superintegrable) $\Om$-anharmonic generalization of ${\cal H}_\SW$ (\ref{dda}) since the former  is  endowed with a single quadratic integral of motion (either  ${\cal I}_{1,\kk}$  or ${\cal I}_{2,\kk}$) and  its flat $\k\to 0$ limit is generally not superintegrable. We remark that in this generalization only $  {\cal I}_{1,\kk}\equiv  {I}_{01}$   `survives'    from the set of three integrals  (\ref{ddb}), meanwhile the two remaining ones are mixed within
   $  {\cal I}_{2,\kk}$, and when $\Om=0$  we get 
  $  {\cal I}_{2,\kk}=I_{02}+\k I_{12}+\k(\la_1+\la_2)$. Note also that the $\k=0$ Euclidean limit is naturally included in (\ref{da}) and (\ref{db}), and provides (\ref{bf}).


\begin{figure}[t]
\setlength{\unitlength}{1mm}
\begin{picture}(140,120)(0,0)
\footnotesize{
\put(15,73){\includegraphics[scale=0.3]{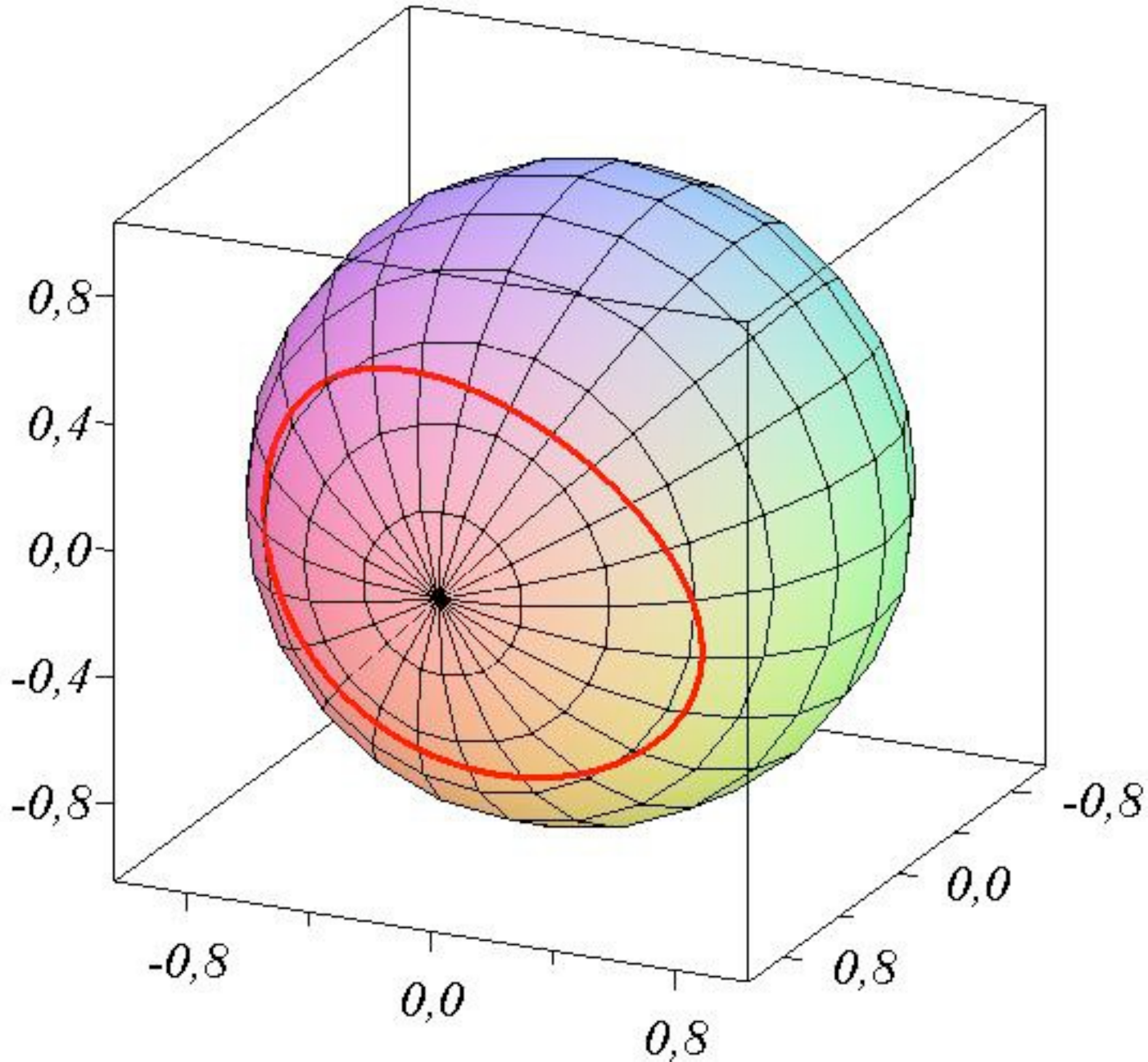}}
\put(33,72){\footnotesize{$x_1$}}
\put(65,78){\footnotesize  $x_0$}
\put(11,98){\footnotesize  $x_2$}
\put(43,67){$(a)$}
\put(85,73){\includegraphics[scale=0.3]{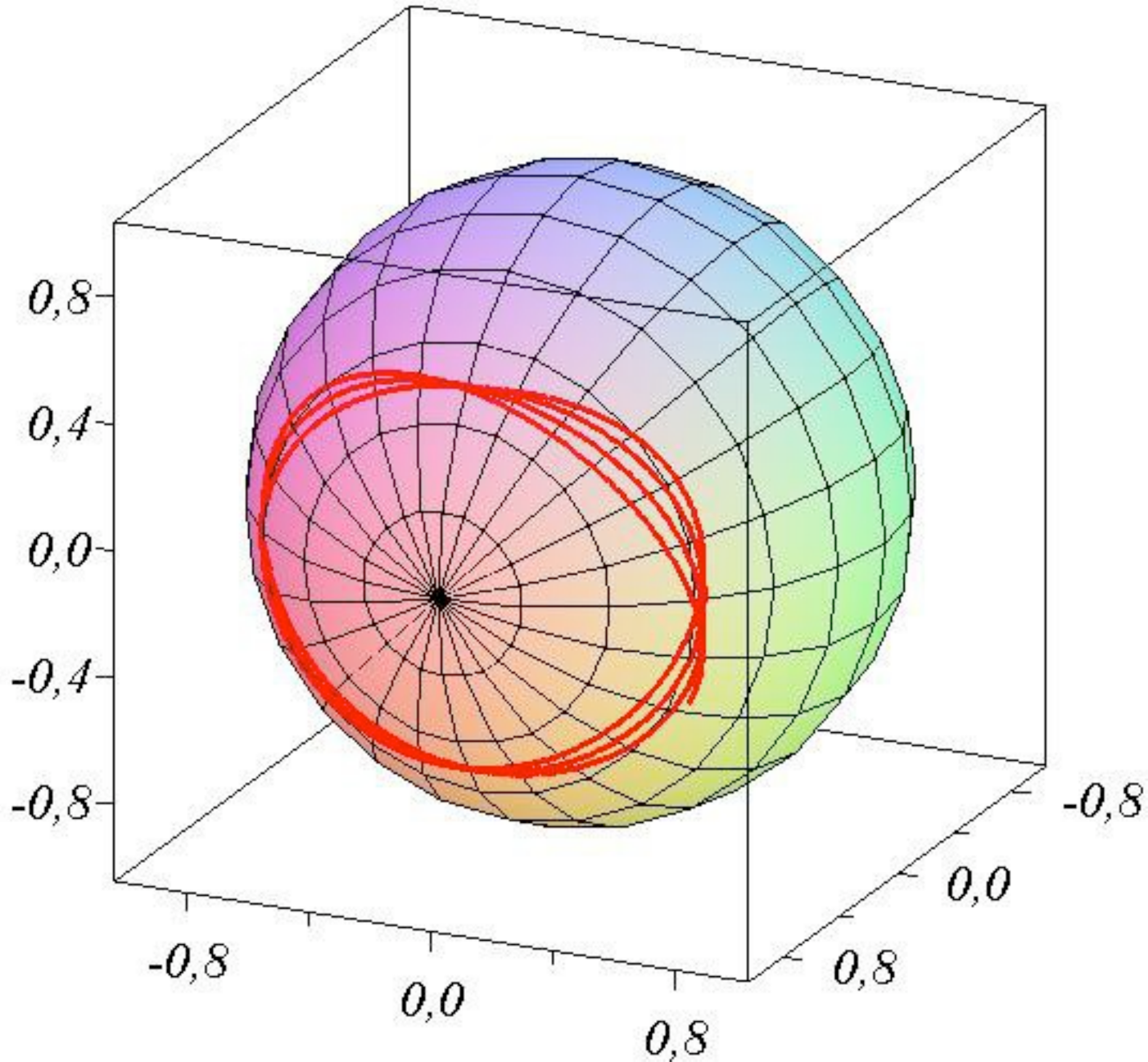}}
\put(103,72){\footnotesize $x_1$}
\put(135,78){\footnotesize  $x_0$}
\put(81,98){\footnotesize $x_2$}
\put(113,67){$(b)$}
\put(15,8){\includegraphics[scale=0.3]{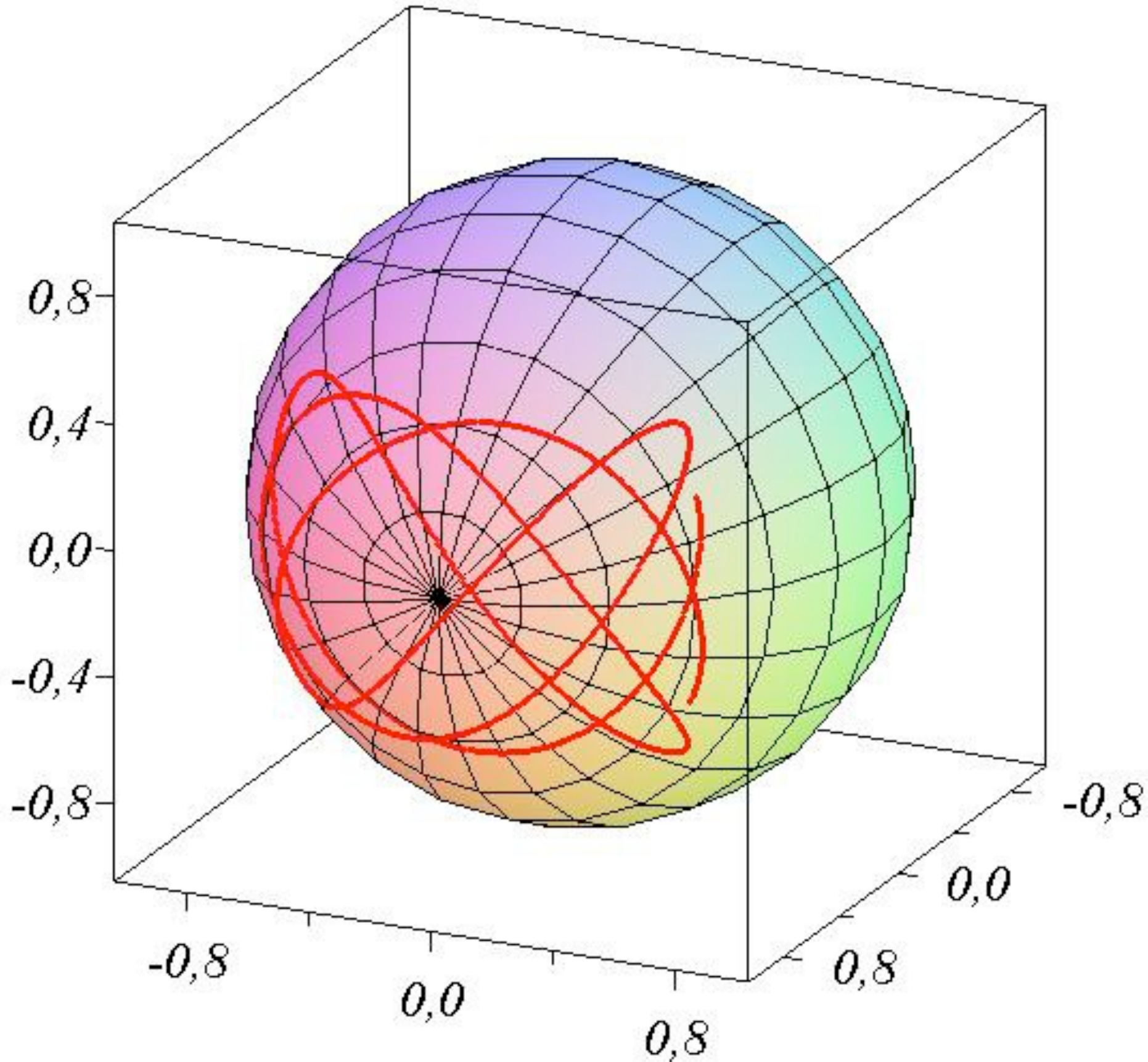}}
\put(33,7){\footnotesize $x_1$}
\put(65,13){\footnotesize  $x_0$}
\put(11,33){\footnotesize $x_2$}
\put(43,2){$(c)$}
\put(85,8){\includegraphics[scale=0.3]{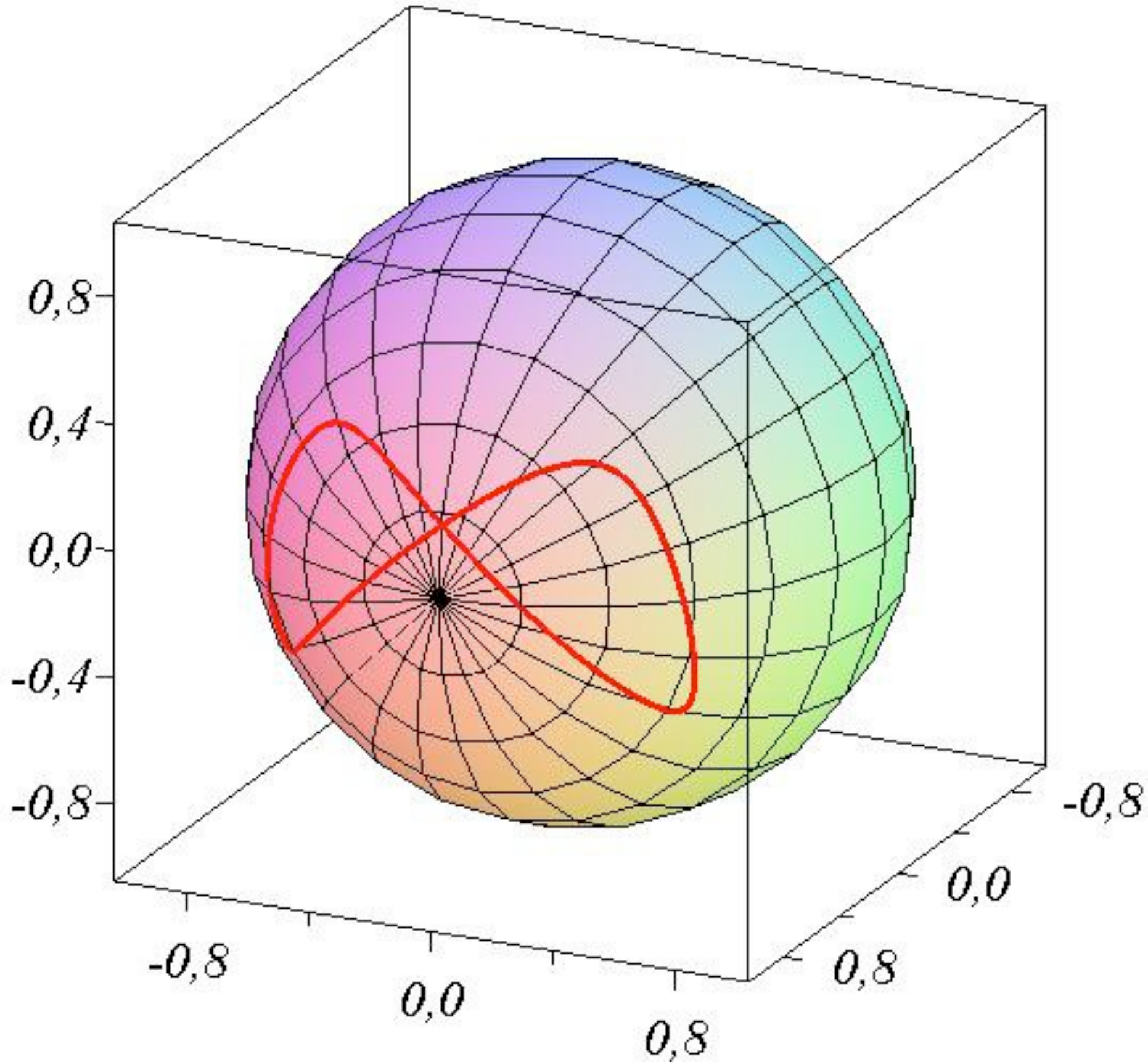}}
\put(103,7){\footnotesize $x_1$}
\put(135,13){\footnotesize $x_0$}
\put(81,33){\footnotesize $x_2$}
\put(113,2){$(d)$}
}
\end{picture}
\caption{  \footnotesize Some trajectories of the Hamiltonians ${\cal H}_\kk$ (\ref{da}) and $\non_\kk$ (\ref{dd})
on the sphere  ${\mathbf S}^2$  with $\k=1$ and without centrifugal terms ($\lambda_1=\lambda_2=0$). They are plotted in  $\mathbb R^{3}$ with   ambient coordinates  $(x_0,\>x)$ such that $x_0^2+x_1^2+x_2^2=1$. Time runs from $t=0$ to $t=8$ for the initial data $q_1=1$, $\dot{q}_1=1$, $q_2=-0.5$, 
$\dot{q}_2=2$:
$(a)$~the superintegrable case of ${\cal H}_\kk$ with $\delta=1$ and   $\Omega=0$ (Higgs oscillator),
$(b)$ the integrable ${\cal H}_\kk$ with $\delta=1$ and  $\Omega=0.2$,
$(c)$  the integrable  ${\cal H}_\kk$ with $\delta=1$ and  $\Omega=3$ and
$(d)$ the superintegrable   $\non_\kk$ with $\delta=1$.     }
\end{figure}
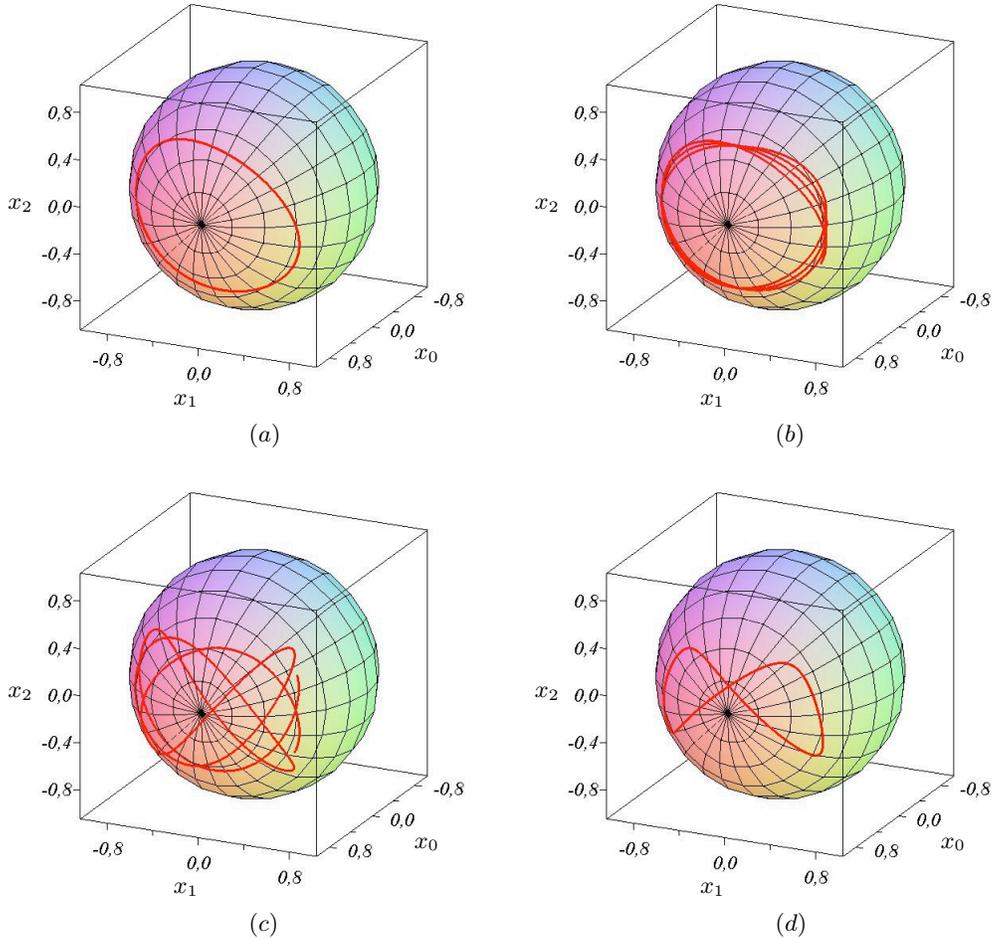



\begin{figure}[t]
\setlength{\unitlength}{1mm}
\begin{picture}(140,115)(0,0)
\footnotesize{
\put(15,73){\includegraphics[scale=0.25]{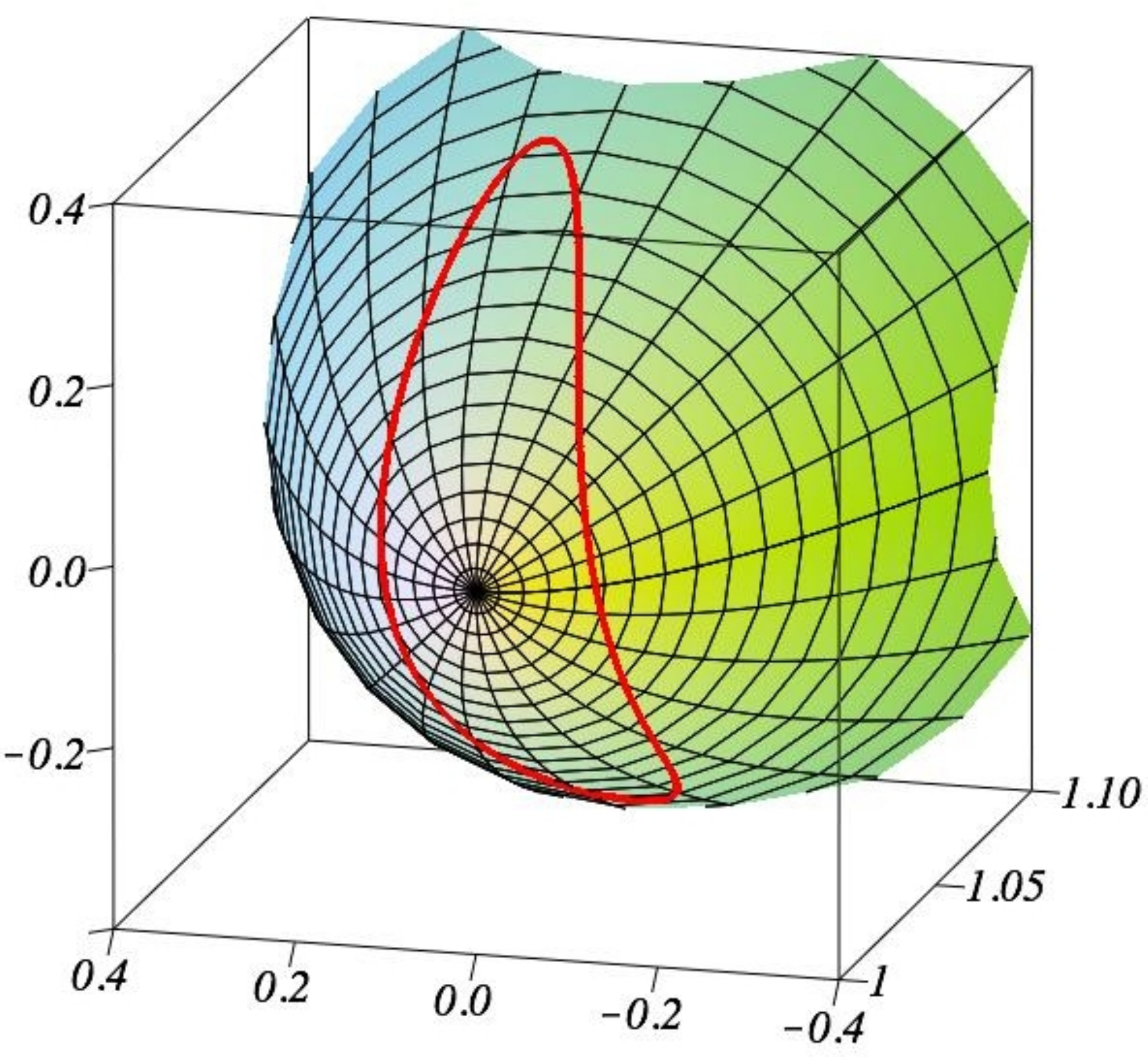}}
\put(33,71){\footnotesize{$x_1$}}
\put(64,77){\footnotesize  $x_0$}
\put(11,95){\footnotesize  $x_2$}
\put(43,66){$(a)$}
\put(85,73){\includegraphics[scale=0.25]{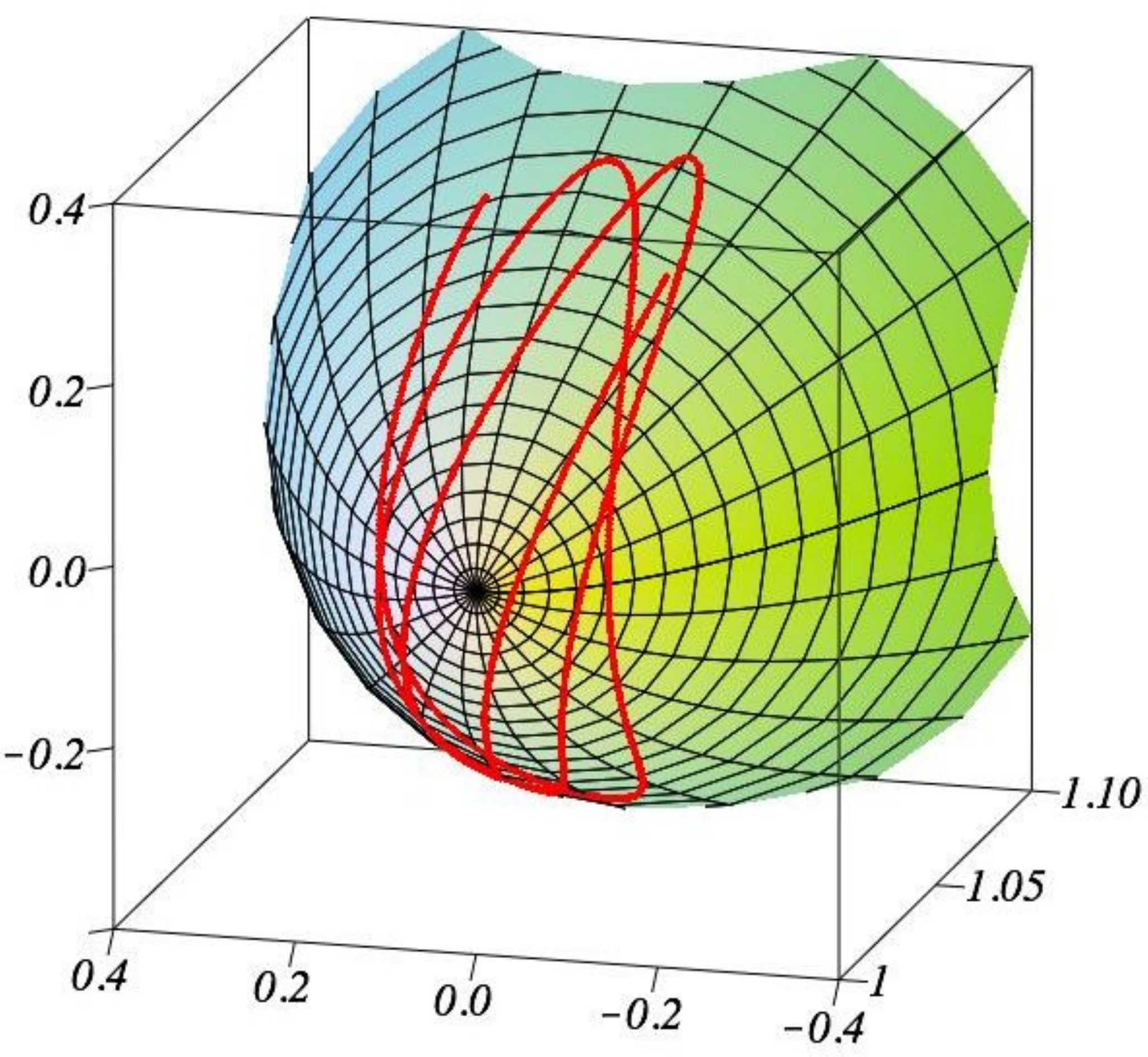}}
\put(103,71){\footnotesize $x_1$}
\put(134,77){\footnotesize  $x_0$}
\put(80,95){ \footnotesize $x_2$}
\put(113,66){$(b)$}
\put(15,11){\includegraphics[scale=0.25]{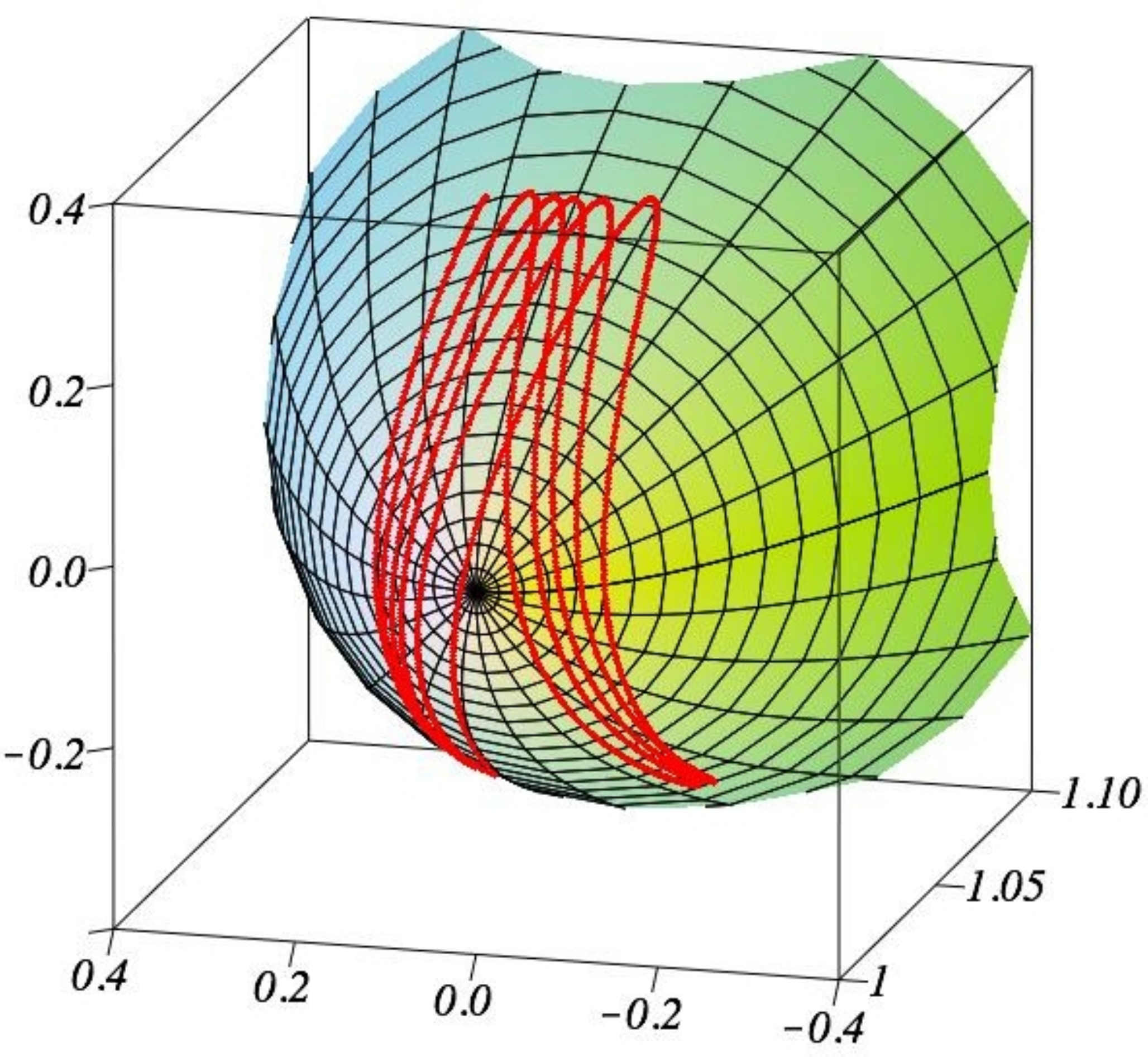}}
\put(33,9){ \footnotesize $x_1$}
\put(64,15){\footnotesize  $x_0$}
\put(10,33){ \footnotesize $x_2$}
\put(43,4){$(c)$}
\put(85,11){\includegraphics[scale=0.25]{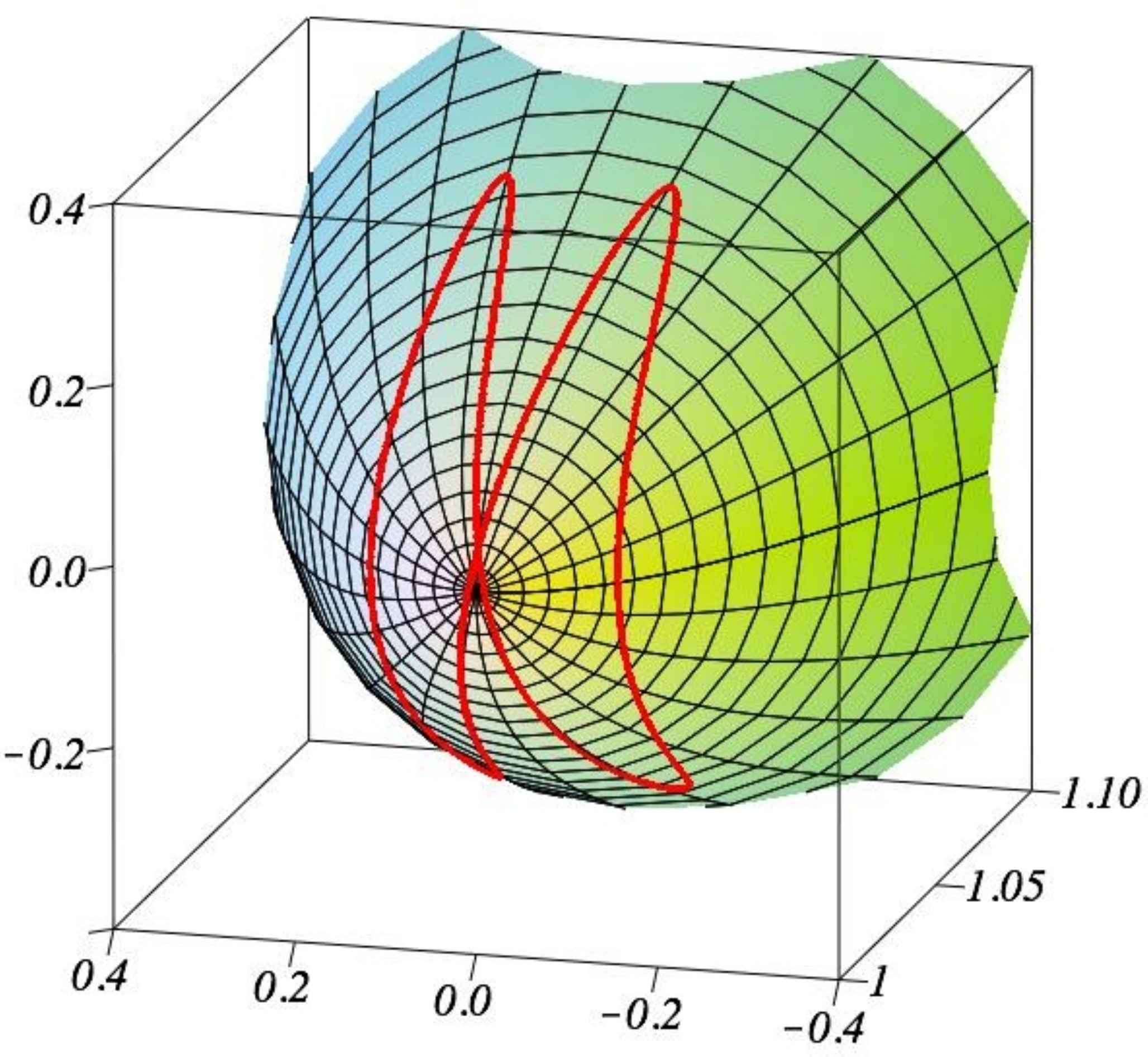}}
\put(103,9){\footnotesize $x_1$}
\put(134,15){\footnotesize $x_0$}
\put(81,33){\footnotesize $x_2$}
\put(113,4){$(d)$}
}
\end{picture}
\caption{\footnotesize Some trajectories of the Hamiltonians ${\cal H}_\kk$ (\ref{da}) and $\non_\kk$ (\ref{dd})
on the hyperboloid  ${\mathbf H}^2$  with $\k=-1$ and without centrifugal terms ($\lambda_1=\lambda_2=0$). They are plotted in  $\mathbb R^{3}$ with   ambient coordinates  $(x_0,\>x)$ such that $x_0^2-x_1^2-x_2^2=1$. Time runs from $t=0$ to $t=12$ with the initial data $q_1=0.1$, $\dot{q}_1=0.1$, $q_2=0.3$, 
$\dot{q}_2=-0.2$:  $(a)$  the superintegrable case of ${\cal H}_\kk$ with $\delta=1$ and   $\Omega=0$ (Higgs oscillator),
$(b)$ the integrable ${\cal H}_\kk$ with $\delta=1$ and  $\Omega=0.2$,
$(c)$~the integrable  ${\cal H}_\kk$ with $\delta=1$ and  $\Omega=3$ and
$(d)$ the superintegrable   $\non_\kk$ with $\delta=1$. }
 \end{figure}


The system defined by proposition 2 can also be written  in terms of Beltrami variables by applying the   transformation (\ref{ck}), and it contains only polynomial and rational expresions 
\bea
&& {\cal H}_\kk=\frac 12 \left(1+\kk \bq^2\right) \left(\bp^2+\kk(\bq\cdot\bp)^2 \right) +\del \bq^2 +\Om\, \frac{q_2^2}{(1+\kk \bq^2)}+
\left(1+\kk \bq^2\right)\left(\frac{\la_1}{q_1^2}+\frac{\la_2}{q_2^2} \right),\nonumber\\[2pt]
 &&{\cal I}_{1,\kk}=\frac{1}{2} J_{01}^2 +\del  q_{1}^{2}  +\frac{\la_1}{q_{1}^{2}}, \label{hhii}\\[2pt]
&&{\cal I}_{2,\kk}=\frac{1}{2} \left( J_{02}^2+\kk J_{12}^2 \right)+\del q_{2}^{2}  +\Om\, \frac{q_2^2}{(1+\kk \bq^2)} +\frac{\la_2}{q_{2}^{2}}+
\kk \bq^2  {\left(\frac{\la_1}{q_1^2}+  \frac{\la_2 } {q_2^2}  \right)},
\nonumber
\eea
where now    $J_{ij}$ are the functions given by (\ref{cd}). In this way, the curved  Higgs oscillator potential, $ \Tk^2_\kk(r)$, becomes   the simple  `flat' expression $\bq^2$. Again,
the Euclidean Hamiltonian  ${\cal H}$ is recovered in Cartesian coordinates, in the form (\ref{ba}) and (\ref{bb}),
by setting   $\k=0$ in (\ref{hhii}).

Numerical solutions for some particular bounded trajectories of  ${\cal H}_\kk$  (\ref{da})  are explicitly illustrated in figures 2 (sphere) and 3 (hyperboloid). They are  obtained by fixing the values of  the parameters $\Om$ and $\del$   (for the sake of simplicity both $\la_i$ are set equal to zero) and by taking the same initial conditions. In this respect, 
it is worth remarking that:

\begin{itemize}

\item The main dynamical footprint of superintegrable systems consists in the fact that all their bounded trajectories are periodic ones. Indeed, when $\Om=\la_1=\la_2=0$ we recover the superintegrable Higgs oscillator, whose periodic trajectories are plotted in figures $2(a)$ and $3(a)$.

\item The   integrable Hamiltonian (\ref{da}), arising by introducing the anharmonic potential with $\Om\ne 0$ (we keep $\la_1=\la_2=0$), corresponds to the anisotropic   Higgs oscillator. In this case the bounded trajectories shown in 
 figures $2(b)$ and $3(b)$  turn out to be nonperiodic. 
 Therefore, the transition from the Higgs oscillator $(\Om=0)$ to the anisotropic one $(\Om\ne 0)$ can be graphically  appreciated by comparing  figures $2(a)$ and  $2(b)$ (see also $3(a)$ and $3(b)$). 

\item We also analyse the particular case of ${\cal H}_\kk$ with  $\Om=3\del$ and $\la_1=\la_2=0$, which is a curved analogue $(\k\ne 0)$ of the  superintegrable Euclidean oscillator 1:2 $(\k=0)$ whose flat trajectory is the    Lissajous 1:2  curve. For the initial conditions here chosen, it is found that the curved bounded trajectories are no longer periodic as it is depicted in figures $2(c)$ and $3(c)$.

\end{itemize}

Therefore, by taking into account these and many other numerical computations that we have performed for different initial conditions and also for other  values of $\Om$ that correspond to commensurate Euclidean oscillators (for instance, with $\Om=8\delta$) we can conjecture that the Hamiltonian ${\cal H}_\kk$  (\ref{da})  seems to be superintegrable only in the $\Om=0$ case.  Nevertheless, we stress that we have not rigurously proven the non-superintegrability of  ${\cal H}_\kk$ for an arbitrary value of  $\Om$. 
Equivalently, we could say that for these commensurate values of $\Om$, the anisotropic Higgs Hamiltonian ${\cal H}_\kk$ presents a superintegrability breaking induced by the non-vanishing curvature of the underlying space.


\sect{The superintegrable curved 1:2 oscillator}

As far as the known superintegrable anisotropic 1:2 oscillator (\ref{bg}) is concerned, its curved counterpart for any value of the curvature $\k$  can be written as follows.

\medskip

\noindent
{\bf Proposition 3.}~\cite{RS} {\em  The Hamiltonian 
\bea
 &&\non_\kk=\frac 12 \left( p_r^2+\frac{p_\te^2}{\Sk^2_\kk(r)} \right) +\del \,\frac{\Sk^2_\kk(r)\cos^2\te}{\left(1-\kk \Sk^2_\kk(r)\cos^2\te\right)}   
  \nonumber\\[2pt]
   && \qquad \qquad  +4\del\, \frac{\Tk^2_\kk(r)\sin^2\te}{\left(1-\kk \Sk^2_\kk(r)\cos^2\te\right)\left(1-\kk \Tk^2_\kk(r)\sin^2\te\right)^2} +\frac{\la_1}{\Sk^2_\kk(r) \cos^2\te} \label{dd}
    \eea
is endowed with three integrals of motion given by
\bea 
   &&\ii_{1,\kk}= \frac{1}{2} \left( J_{01}^2+\kk J_{12}^2 \right)+\del\, \frac{ \Tk^2_\kk(r)\cos^2\te {\left(1+\kk \Tk^2_\kk(r)\sin^2\te\right)} } {\left(1-\kk \Tk^2_\kk(r)\sin^2\te\right)^2} 
+\frac{\la_1}{\Sk^2_\kk(r) \cos^2\te}, \label{df}\\[2pt]
     &&\ii_{2,\kk}=\frac{1}{2} J_{02}^2+4 \del\, \frac{ \Tk^2_\kk(r)\sin^2\te} {\left(1-\kk \Tk^2_\kk(r)\sin^2\te\right)^2} 
   ,   \nonumber     \\[2pt]
     &&\ele_{\kk}= J_{01}J_{12}+2\del\, \frac{ \Tk^3_\kk(r)\cos^2\te\sin\te} {\left(1-\kk \Tk^2_\kk(r)\sin^2\te\right)^2} 
   -2\la_1\, \frac{\tan\te}{\Tk_\kk(r)\cos\te} ,\nonumber
 \eea
where $J_{ij}$ are the functions  (\ref{ch}) and   $\non_\kk=\ii_{1,\kk}+\ii_{2,\kk}$.  Both sets  $(\non_\kk , \ii_{1,\kk} ,\ele_{\kk}  )$ and $(\non_\kk, \ii_{2,\kk} , \ele_{\kk} )$ are formed by three functionally independent functions.
 }
\medskip

Once again, this system can be more easily expressed by introducing  Beltrami variables through (\ref{ck}). In this way we find that
 \bea
&&\non_\kk=\frac 12 \left(1+\kk \bq^2\right) \left(\bp^2+\kk(\bq\cdot\bp)^2 \right) +\frac{\del \, q_1^2}{(1+\kk q_2^2)}
+\frac{4\del\,   (1+\kk \bq^2)q_2^2}{ (1+\kk q_2^2)(1-\kk q_2^2)^2  }  +\la_1\,\frac{(1+\kk \bq^2)}{q_1^2} ,\nonumber\\[2pt]
&& \ii_{1,\kk}= \frac{1}{2} \left( J_{01}^2+\kk J_{12}^2 \right)   + \del\, \frac{q_1^2(1+\kk q_2^2) }{(1-\kk q_2^2)^2}  +\la_1\,\frac{(1+\kk \bq^2)}{q_1^2} , \label{de} \\[2pt]
&&
\ii_{2,\kk}=\frac{1}{2} J_{02}^2 +4\del\,   \frac{q_2^2}{(1-\kk q_2^2)^2 }  , \qquad
 \ele_{\kk}=  J_{01} J_{12}  +2\del\,   \frac{q_1^2q_2}{(1-\kk q_2^2)^2 }-2\la_1\,  \frac{q_2}{q_1^2} ,
 \nonumber
\eea
where      $J_{ij}$ are now the functions   (\ref{cd}). In this variables,  the Euclidean case with $\k=0$ leads just to the equations  (\ref{bd})
defining $\non$ in Cartesian coordinates.


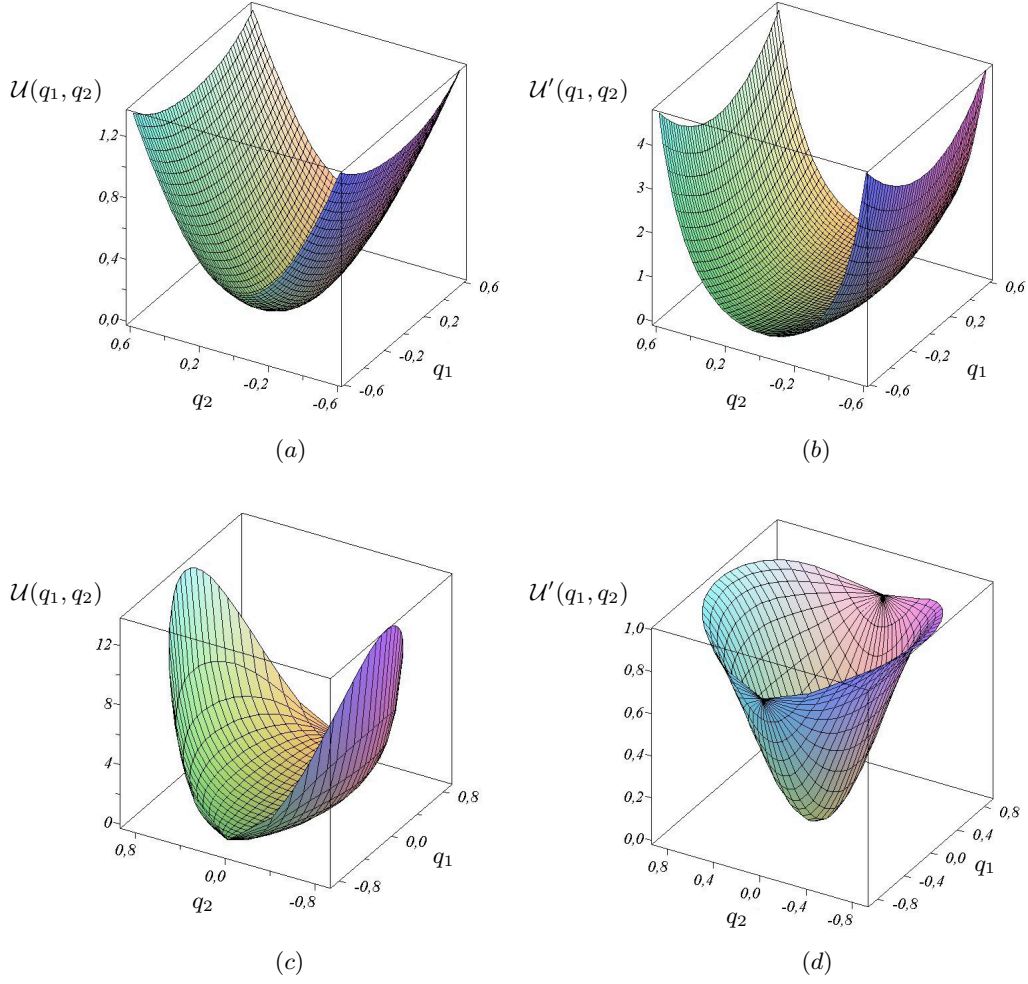
\begin{figure}[t]
\setlength{\unitlength}{1mm}
\begin{picture}(140,132)(0,0)
\footnotesize{
\put(15,69){\includegraphics[scale=0.25]{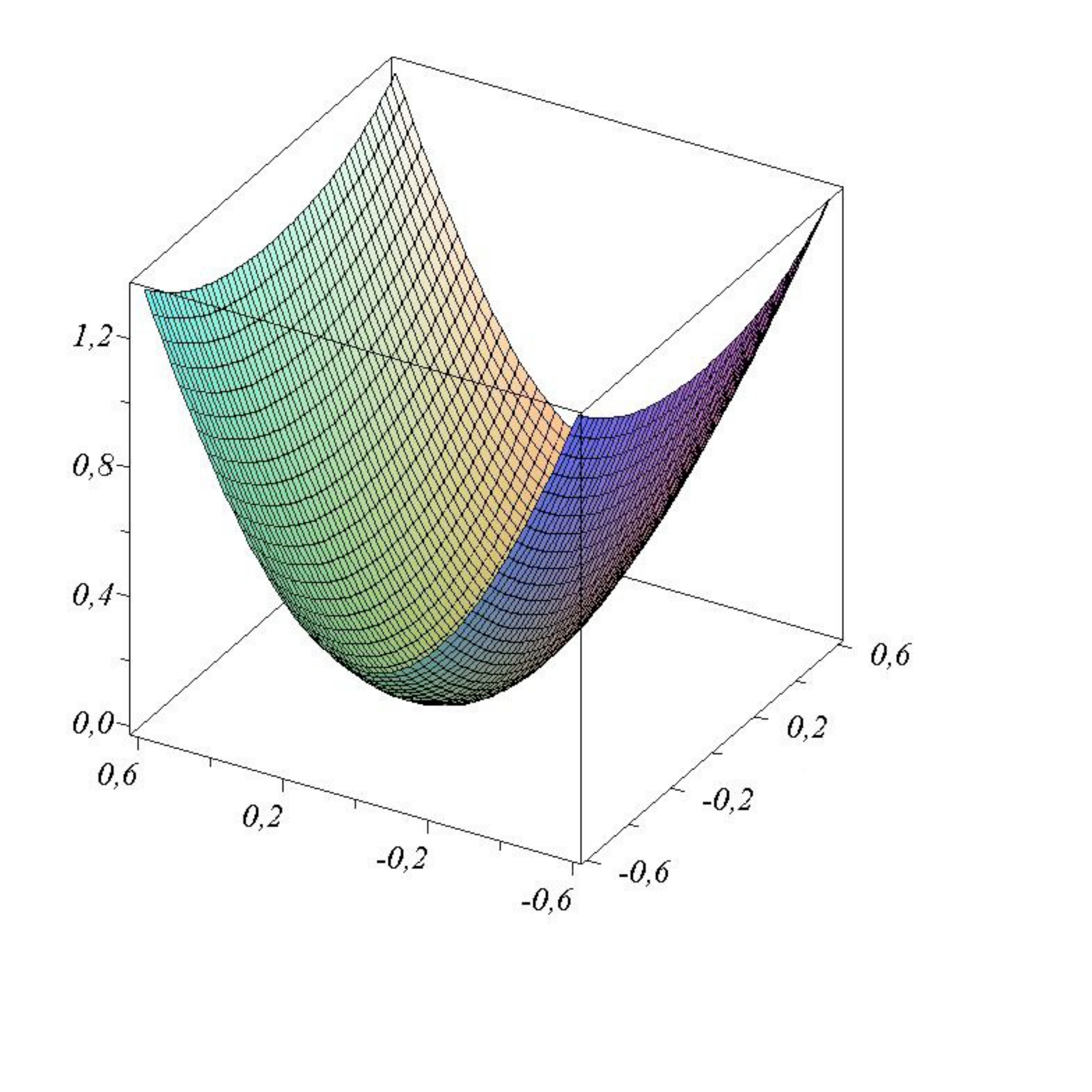}}
\put(32,81){\footnotesize $q_2$}
\put(64,85){\footnotesize $q_1$}
\put(8,122){\footnotesize ${\cal U}(q_1,q_2)$}
\put(43,74){$(a)$}
\put(85,69){\includegraphics[scale=0.25]{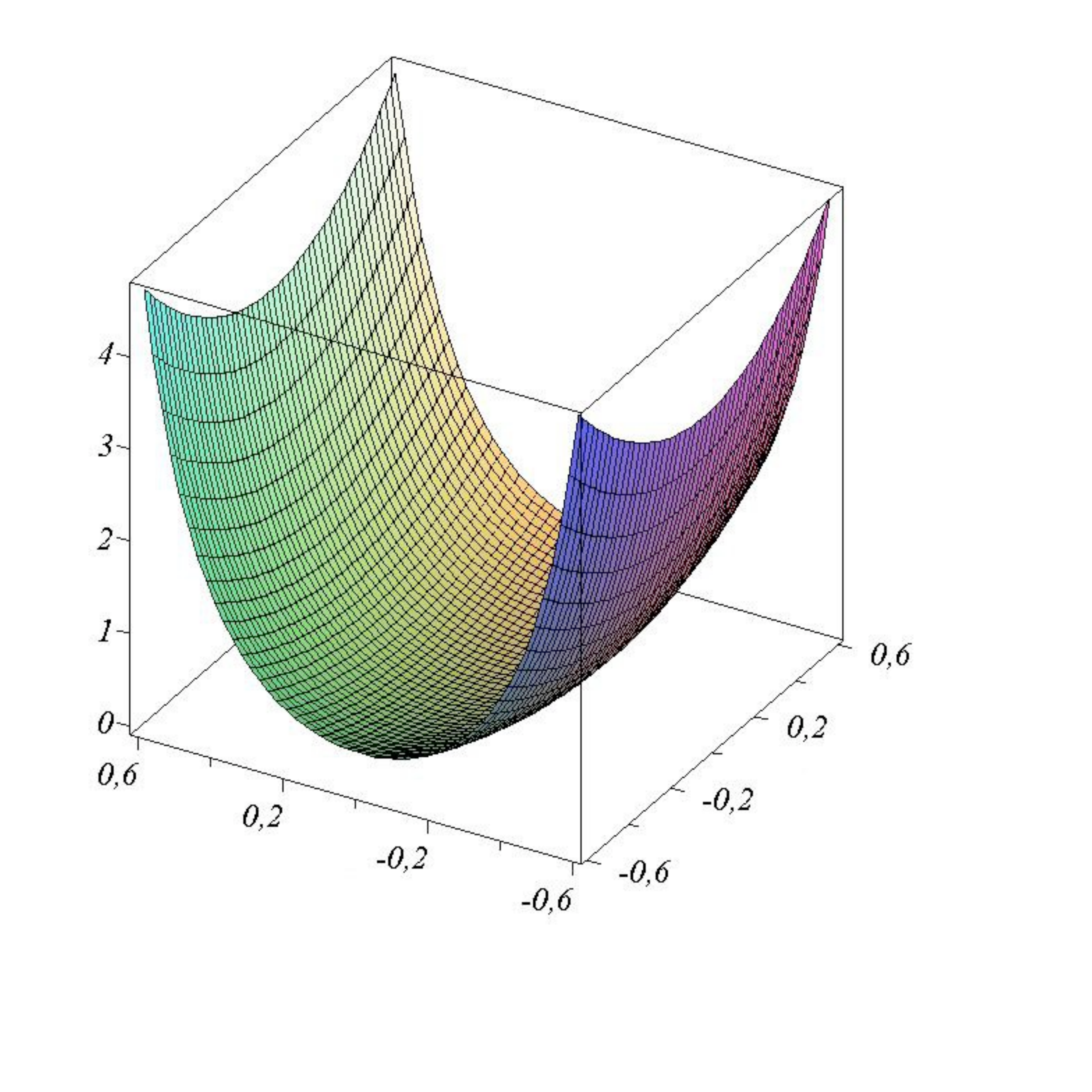}}
\put(103,81){\footnotesize $q_2$}
\put(135,85){\footnotesize $q_1$}
\put(77,122){\footnotesize ${\cal U}'(q_1,q_2)$}
\put(113,74){$(b)$}
\put(11,0){\includegraphics[scale=0.26]{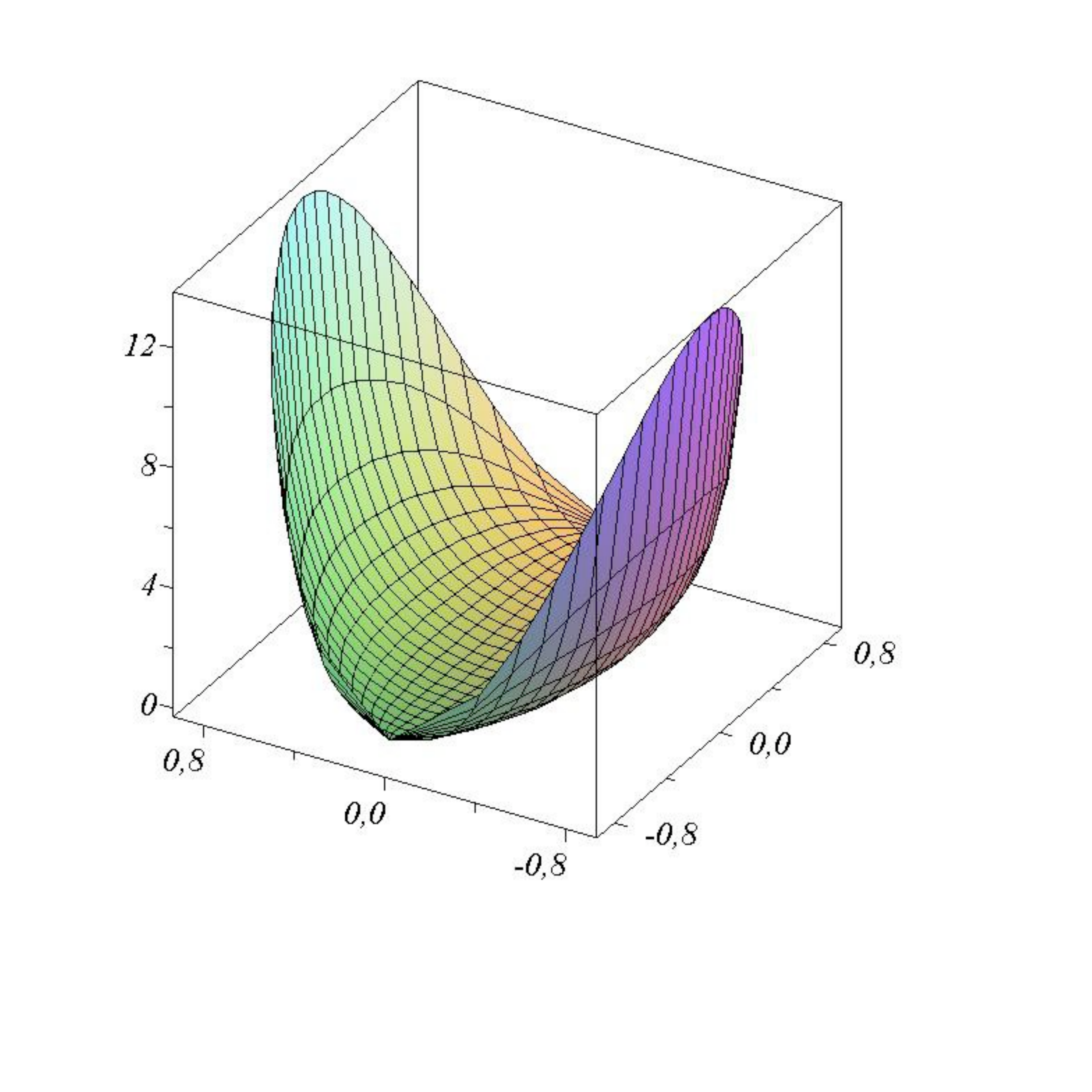}}
\put(32,14){\footnotesize $q_2$}
\put(64,20){\footnotesize $q_1$}
\put(8,55){\footnotesize ${\cal U}(q_1,q_2)$}
\put(43,6){$(c)$}
\put(85,0){\includegraphics[scale=0.25]{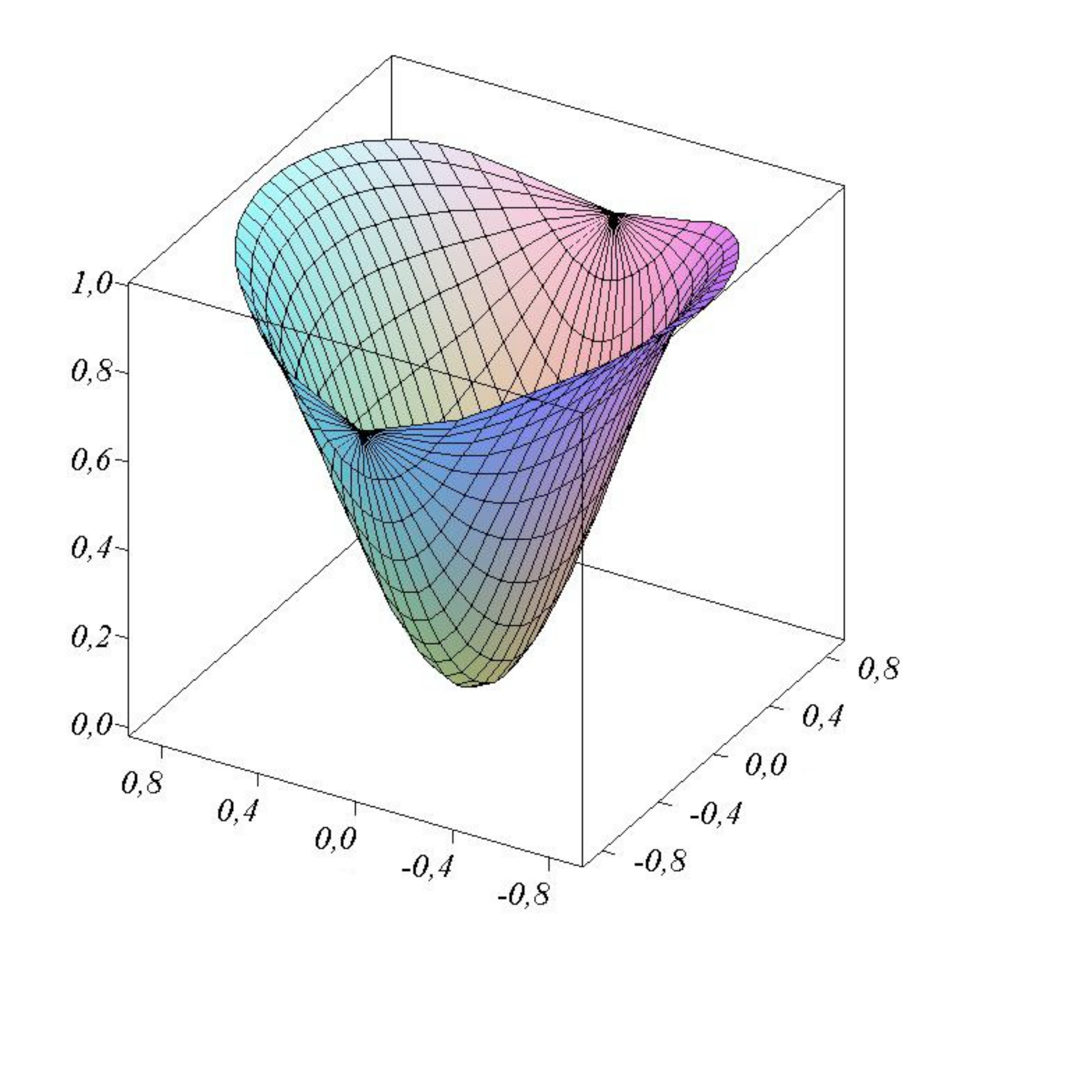}}
\put(103,12){\footnotesize $q_2$}
\put(136,19){\footnotesize $q_1$}
\put(77,55){\footnotesize ${\cal U}'(q_1,q_2)$}
\put(113,6){$(d)$}
}
\end{picture}
\caption{ \footnotesize The potentials ${\cal U}$ (\ref{hhii})  and ${\cal U}'$ (\ref{de})  (without centrifugal terms $\lambda_1=\la_2=0$)  in Beltrami coordinates   on the sphere ${\mathbf S}^2$ with $\k=1$ and on the hyperbolic plane  ${\mathbf H}^2$  with $\k=-1$:  $(a)$  the integrable    ${\cal U}$ on  ${\mathbf S}^2$ for $\delta=1$ and $\Omega=3$,
$(b)$ the superintegrable    ${\cal U}'$ on  ${\mathbf S}^2$ for $\delta=1$, 
$(c)$ the integrable  ${\cal U}$ on  ${\mathbf H}^2$   for $\delta=1$ and $\Omega=3$ and 
$(d)$~the superintegrable  ${\cal U}'$ on  ${\mathbf H}^2$ for $\delta=1$.}
\end{figure}


The numerical integration of the trajectories for the superintegrable Hamiltonian $\non_\kk$ (\ref{dd}) gives rise to the Lissajous 1:2 curves on the sphere and the hyperboloid which are shown in figures 2$(d)$ and 3$(d)$, respectively. They are worthy to be compared with figures 2$(c)$ and 3$(c)$.
Hence, in contrast with the integrable Hamiltonian ${\cal H}_\kk$ (\ref{da}) with $\Om=3\del$, 
the system  $\non_\kk$  can be regarded as the appropriate curved analogue ($\k\ne 0$) of  the anisotropic Euclidean oscillator 1:2 ($\k=0$). We recall that in the Euclidean plane  $\non$ (\ref{bd}) is just the particular case of  ${\cal H}$ (\ref{ba}) with  $\Om=3\del$ and $\la_2=0$, but this is no longer true when the curvature is turned on. Consequently, we have shown that the Euclidean 1:2 oscillator admits (at least) two different  integrable curved generalizations, since $\non_\kk$  (the superintegrable one) cannot be recovered as a particular case of ${\cal H}_\kk$   with $\Om=3\del$ and $\la_2=0$.

 The difference between these two systems can also be highlighted by representing the corresponding potentials. 
We write
${\cal H}_\kk={\cal T}+{\cal U}$  (\ref{hhii}) and $\non_\kk={\cal T}+{\cal U}'$  (\ref{de}), with ${\cal T}$ given by (\ref{ce}), and we represent    both potentials  ${\cal U}$ (with $\Om=3\del$) and ${\cal U}'$ in Beltrami coordinates in figure~4.
Notice  that the corresponding trajectories for the potentials shown in figures 4$(a)$, 4$(b)$, 4$(c)$ and 4$(d)$ are  drawn in figures 2$(c)$, 2$(d)$, 3$(c)$ and 3$(d)$, respectively.

On the other hand,  the differences between  these two curved anisotropic  potentials can also  be   enhanced by rewritting them in terms of the ambient coordinates  $(x_0,\>x)$ introduced in section 2, namely
\bea
&&{\cal U}=   \del \,\frac{\>x^2}{(1-\k \>x^2)} +\Om\, x_2^2 +\frac{\la_1}{x_1^2} +\frac{\la_2}{x_2^2} ,\nonumber \\[2pt]
&&  {\cal U}'=  \del\, \frac{x_1^2}{(1-\k x_1^2)}+4\del\, \frac{x_0^2x_2^2}{(x_0^2+\k x_2^2)(x_0^2-\k x_2^2)^2}+\frac{\la_1}{x_1^2} ,\label{dddff}
\eea
and recall that $x_0\to 1$ when $\k\to 0$. In this form, these potentials can  be more easily  compared with the proposals for integrable anharmonic oscillators on the spheres and hyperbolic spaces that have been previously presented  in~\cite{kalnins, Saksida}, thus concluding that the anisotropic Higgs oscillator potential $\cal U$ defines a new integrable system. 
In fact, the potentials studied in~\cite{kalnins, Saksida}  are just polynomials when written  in ambient coordinates. For instance,  the integrable potential on the $n$-dimensional sphere considered in~\cite{Saksida} reads
  $$
  {\cal V}=\sum_{i=0}^n a_i^2 x_i^2-\left( \sum_{i=0}^n a_i x_i^2 \right)^2 ,
  $$
  where $a_i$ are arbitrary real constants and $x_i$ are $(n+1)$  ambient coordinates such that  $\sum_{i=0}^n x_i^2=1$. The systems worked out in~\cite{kalnins} have a similar form. Evidently, they are quite  different from~(\ref{dddff}).


\sect{Curved  `centrifugal'  terms: geometry and dynamics}

It is worth recalling that the  SW system on ${\mathbf S}^2$ was
interpreted in~\cite{ran,ran1,ran2} as a superposition of three spherical oscillators, and this result has been extended to higher dimensions and to other spaces with constant curvature   in~\cite{int,kiev,CRMVulpi}. The crucial point is to take into account  that the  Higgs oscillator potential given by  $\Tk^2_\kk(r)$,   is  a {\em central}  harmonic oscillator,  whose center is located
at the origin $O$,   and $r$  is the   distance  between  $Q$  and $O$ measured along 
the geodesic $l$ that joins both points (see section 2).
 Hence if, instead of $O$, we take a generic point $Q_\rho$ such that 
 $\rho$ is the distance between  $Q$ and $Q_\rho$ measured along 
the geodesic $l_\rho$ that joins both points, then $\Tk^2_\kk(\rho)$ can be interpreted as  a {\em noncentral} harmonic oscillator potential with center at $Q_\rho$.
With these ideas in mind, let us start with the potentials ${\cal U}$ and ${\cal U}'$ written in geodesic polar coordinates (\ref{da}) and (\ref{dd})    in order to  search  for terms of the type $\Tk^2_\kk(\rho)$.
 
 Let us consider the two base geodesics $l_1$  and $l_2$ orthogonal at the origin $O$ and the point $Q(r,\te)$. Let $Q_1$  (resp.~$Q_2$)  be the intersection point of $l_2$ (resp.~$l_1$) with its orthogonal   geodesic $l'_1$ (resp.~$l'_2$),     $x= Q Q_1$  the distance  measured along  $l'_1$
and    $y= Q Q_2$  the  one measured along   $l'_2$  as depicted in figure 1. 
Next we apply  the sine theorem~\cite{trigo} on the orthogonal triangles $OQQ_1$ (with  external 
angle $\te$) and $OQQ_{2}$ (with inner
angle $\te$) which yields
$$
OQQ_1:\quad \Sk_{\kk}(x)=\Sk_{\kk}(r)\cos\te  \,;\qquad
OQQ_{2}:\quad    \Sk_{\kk}(y)=\Sk_{\kk}(r)\sin\te.
$$

Therefore the    integrable potential ${\cal U}$ (\ref{da})  can be expressed as
\begin{equation}
{\cal U} 
=\del\Tk^2_{\kk}(r)+ \Om \Sk^2_\kk (y) + \frac{\la_1}{\Sk_{\kk}^2 (x) }+
\frac{\la_2}{\Sk_{\kk}^2(y)}   ,
\label{gf}
\end{equation}
 which provides a unified description of this potential on the three Riemannian spaces:
\begin{itemize}
\itemsep=0pt
\item  The $\del$-term   is a central harmonic oscillator  with center at the origin $O$.

\item The $\Om$-term is the anharmonic potential which only depends on the geodesic distance $y$ to the point $Q_2$, which is not a fixed center since it is moving together with the particle located at $Q$.

\item  The  $\la_i$-terms $(i=1,2)$ are two   `centrifugal  barriers'.
 \end{itemize}

Likewise, the   superintegrable potential ${\cal U}'$ (\ref{dd}) can be rewritten as
\begin{equation}
{\cal U}' 
=\del\Tk^2_{\kk}(x)+ 4\del \Tk^2_\kk (y)\left(\frac{   \Ck^2_\kk (r) \Ck^2_\kk (y)  }{  \Ck^2_\kk(x) \left(\Ck^2_\kk (r) -\kk \Sk^2_\kk (y)  \right)^2}  \right)+ \frac{\la_1}{\Sk_{\kk}^2 (x) }    ,
\label{gg}
\end{equation}
showing that this potential is not a simple addition of some  anharmonic potential (only depending on the variable $y$)  to the {\em central} harmonic oscillator; its interpretation is as 
follows:
 \begin{itemize}
\itemsep=0pt
\item  The first term   is a  noncentral harmonic oscillator  with center at the point $Q_1$ (which is, moreover,  not fixed).

\item The second one is not any kind of curved harmonic oscillator, although it could be regarded as  some kind of   `deformation' of a  noncentral
oscillator  with center at   $Q_2$.

\item  The  $\la_1$-term is  a   `centrifugal  barrier'.
 \end{itemize}
 
 It is worth stressing that the Euclidean limit  is well defined in both (\ref{gf}) and (\ref{gg}), namely
\be
\k\to 0:\qquad  {\cal U}=\del r^2+ \Om  y^2 + \frac{\la_1}{x^2 }+
\frac{\la_2}{y^2} ,\qquad  {\cal U}' =\del x^2+ 4\del  y^2 + \frac{\la_1}{x^2 }  ,
\label{gghh}
\ee
 where now the geodesic distances $x$ and $y$ become  Cartesian coordinates (that is, $q_1$ and $q_2$ appearing in expressions (\ref{ba}) and (\ref{bd})).

Furthermore,  on the sphere ${\mathbf S}^2$ the potential $\cal U$ admits a second interpretation,  which requires to  rewrite the   `centrifugal barriers' as noncentral harmonic oscillators.  Explicitly, let us consider the potential ${\cal U}$ in the form (\ref{gf}), set $\kk=+1$ for simplicity  and 
let  $O_i$ be the {\em fixed} point placed along the   geodesic $l_i$ $(i=1,2)$ which is at a distance 
$\frac{\pi}2$ from   the origin $O$.  Notice that in ambient coordinates $(x_0,\>x)$ we have that
$O=(1,0,0)$, $O_1=(0,1,0)$ and $O_2=(0,0,1)$.
We introduce the  
   distance $r_i$ between $Q$ and $O_i$ measured along  $l'_i$ such that the three points 
    $\{O_iQQ_{i} \}$  lie on the same geodesic $l'_i$ as it is shown in figure 5$(a)$ in the ambient space ${\mathbb R}^3$. Thus we obtain that 
$$
r_1+x=r_2+y =\frac{\pi}2 ,\qquad x_0=\cos r,\qquad x_1= \sin x =\cos r_1,\qquad x_2=\sin y =\cos r_2 .
$$
Then the potential  (\ref{gf}) on  ${\mathbf S}^2$ can  be expressed as 
 \bea
&&{\cal U} 
=\del\tan^2 r+ \Om \sin^2 y + \frac{\la_1}{\sin^2 x }+
\frac{\la_2}{\sin^2 y}   \nonumber\\[2pt]
&&\quad =\del\tan^2 r+ \frac{\Om }{(1+ \tan^2 r_2 )}+  {\la_1}\tan^2 r_1+
 {\la_2}\tan^2 r_2  +\la_1+\la_2 .
\label{gh}
\eea
Hence this corresponds to the superposition of the  central (Higgs) oscillator with center at $O$
with an anharmonic potential depending on the distance $r_2$ to $O_2$ and with two  more noncentral oscillators determined by the geodesic distances $r_i$ from the centers 
 placed at  the fixed points $O_i$. Notice that the flat limit $\k=0$ can only be performed  in the former expression of (\ref{gh}) providing (\ref{gghh}) with the two Euclidean centrifugal barriers, since in the latter the distances  $r_i\to \infty$ (that is, the centers $O_i$ go to infinity) so that the terms $\tan^2 r_i$ become indeterminate.


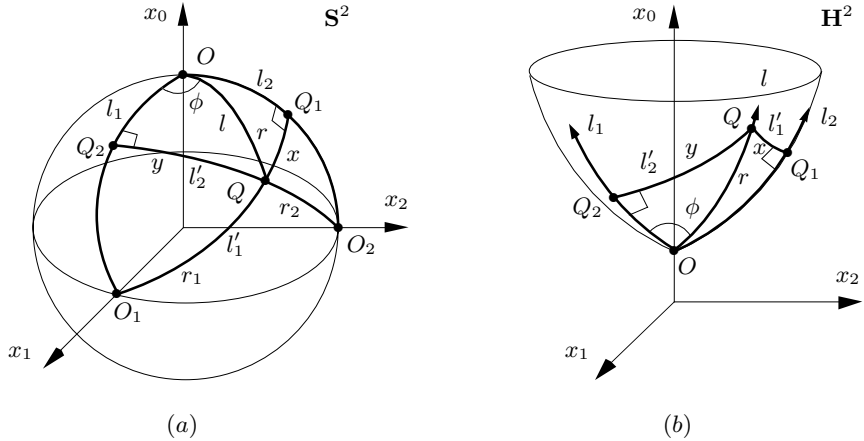
\begin{figure}[t]
\setlength{\unitlength}{1mm}
\begin{picture}(120,60)(0,0)
\footnotesize{
\put(14,8){\includegraphics[scale=1]{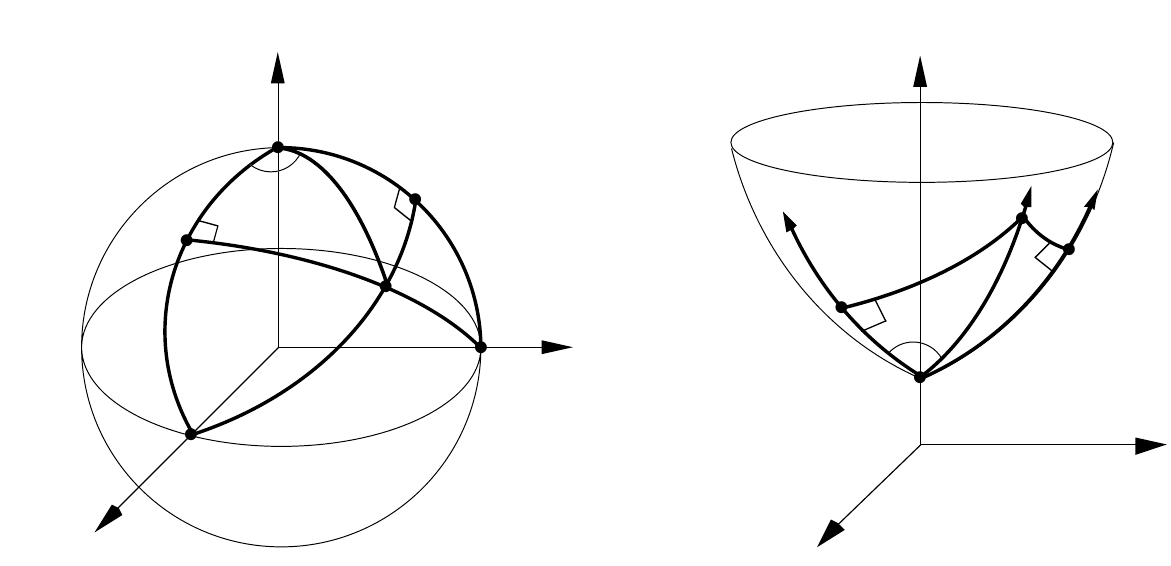}}
\put(37,58){$x_0$}
\put(19,13){{$x_1$}}
\put(69,33){$x_2$}
 \put(44,52){$O$}
\put(33,18){{$O_1$}}
\put(64,27){$O_2$}
 \put(43,36){$l'_2$}
 \put(38,38){$y$}
\put(55,32){$r_2$}
 \put(48,34){$Q$}
 \put(56,39){$x$}
\put(42,23){$r_1$}
\put(48,27){$l'_1$}
 \put(57,46){$Q_1$}
 \put(28,40){$Q_2$}
  \put(52,49){$l_2$}
 \put(32,45){$l_1$}
\put(43,46){$\phi$}
 \put(47,43){$l$}
 \put(52,42){$r$}
  \put(102,58){$x_0$}
\put(93,13){{$x_1$}}
\put(129,23){$x_2$}
\put(119,49){$l$}
\put(114,44){$Q$}
\put(96,43){$l_1$}
\put(127,44){$l_2$}
\put(103,38){$l'_2$}
\put(108,24){$O$}
\put(123,37){$Q_1$}
\put(94,32){$Q_2$}
\put(116,36){$r$}
\put(109,32){$\phi$}
\put(118,40){$x$}
\put(120,43){$l'_1$}
\put(109,40){$y$}
\put(61,57){${\bf S}^2$}
\put(127,57){${\bf H}^2$}
\put(40,3){$(a)$}
\put(106,3){$(b)$}
}
\end{picture}
\caption{\footnotesize Geodesic distances involved in the geometric description of the integrable anisotropic Higgs oscillator with two Rosochatius terms  ${\cal U}$: $(a)$ on the sphere for the expression (\ref{gh})  
and $(b)$ on the hyperboloid in the form  (\ref{gi}).  }
\end{figure}


We remark that in the hyperbolic plane with $\kk<0$    a similar description in terms of oscillators  is, in principle, precluded.  The analogous points to the previous centers
$O_i$   would be beyond infinity as it can be seen in   figure 5$(b)$, that is, beyond the  `proper'  hyperbolic space ${\bf H}^2$.  In particular, if we set $\k=-1$ we find that
   $$
x_0=\cosh r,\qquad x_1= \sinh x,\qquad x_2=\sinh y  ,
$$
  and the potential  (\ref{gf}) on  ${\mathbf H}^2$ reads as  
 \bea
&&{\cal U} 
=\del\tanh^2 r+ \Om \sinh^2 y + \frac{\la_1}{\sinh^2 x }+
\frac{\la_2}{\sinh^2 y}   \nonumber\\[2pt]
&&\quad =\del\tanh^2 r+ \Om \sinh^2 y + \frac{\la_1}{\tanh^2 x }+
\frac{\la_2}{\tanh^2 y}   + \la_1+\la_2 .
\label{gi}
\eea
Hence the $\la_i$-terms only admit an interpretation as centrifugal barriers with Euclidean limit given by (\ref{gghh}).
Nevertheless, an interpretation of the $\la_i$-potentials as noncentral oscillators can be established if one considers 
 the {\em complete} hyperbolic space. In that case, the centers $O_i$   would be located   in the   `ideal'  (exterior) region of ${\bf H}^2$ so beyond the  `actual' ${\bf H}^2$.


\begin{figure}[t]
\setlength{\unitlength}{1mm}
\begin{picture}(140,120)(0,0)
\footnotesize{
\put(14,73){\includegraphics[scale=0.26]{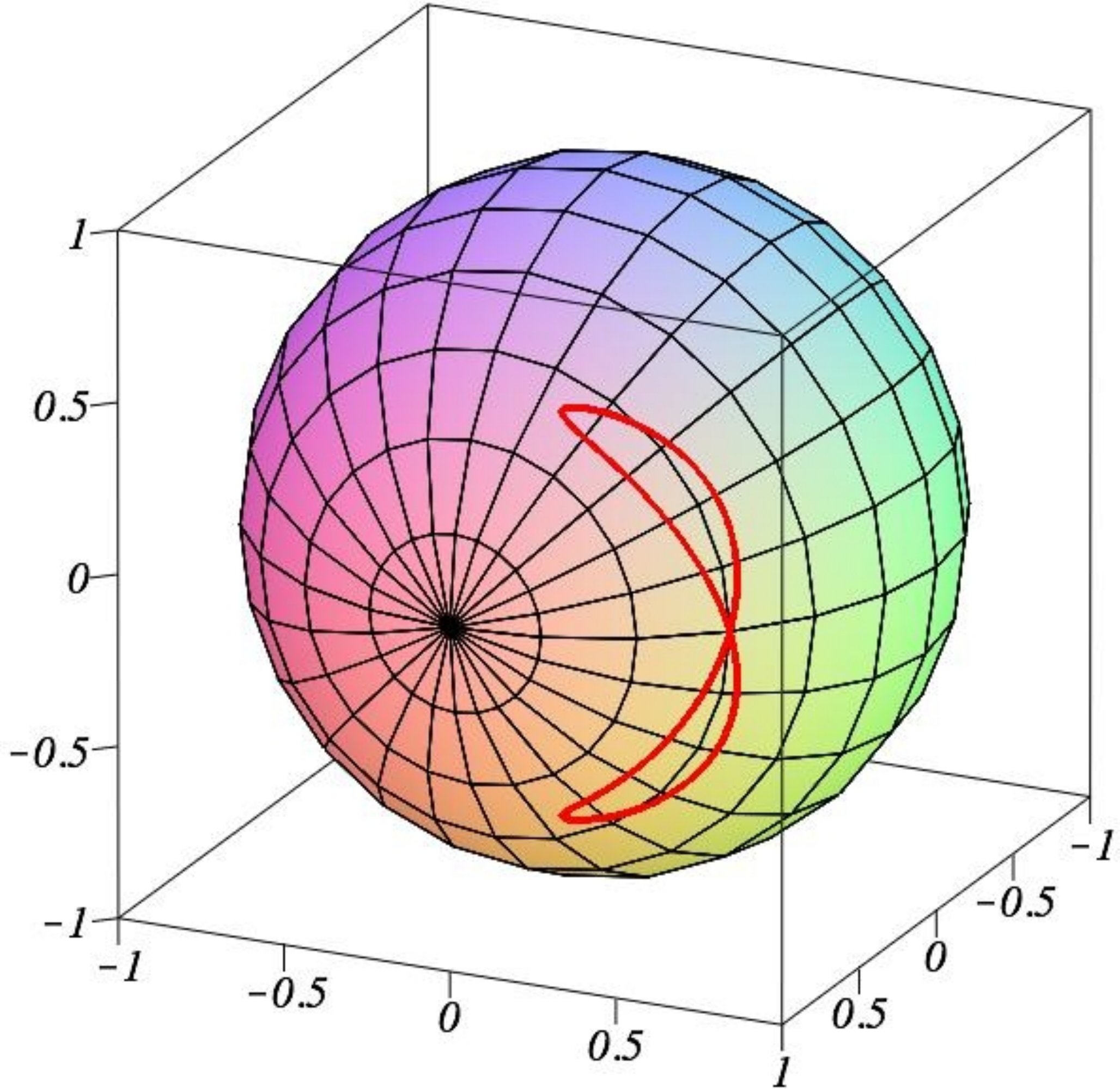}}
\put(33,72){\footnotesize{$x_1$}}
\put(62,76){\footnotesize  $x_0$}
\put(12,98){\footnotesize  $x_2$}
\put(43,67){$(a)$}
\put(84,73){\includegraphics[scale=0.26]{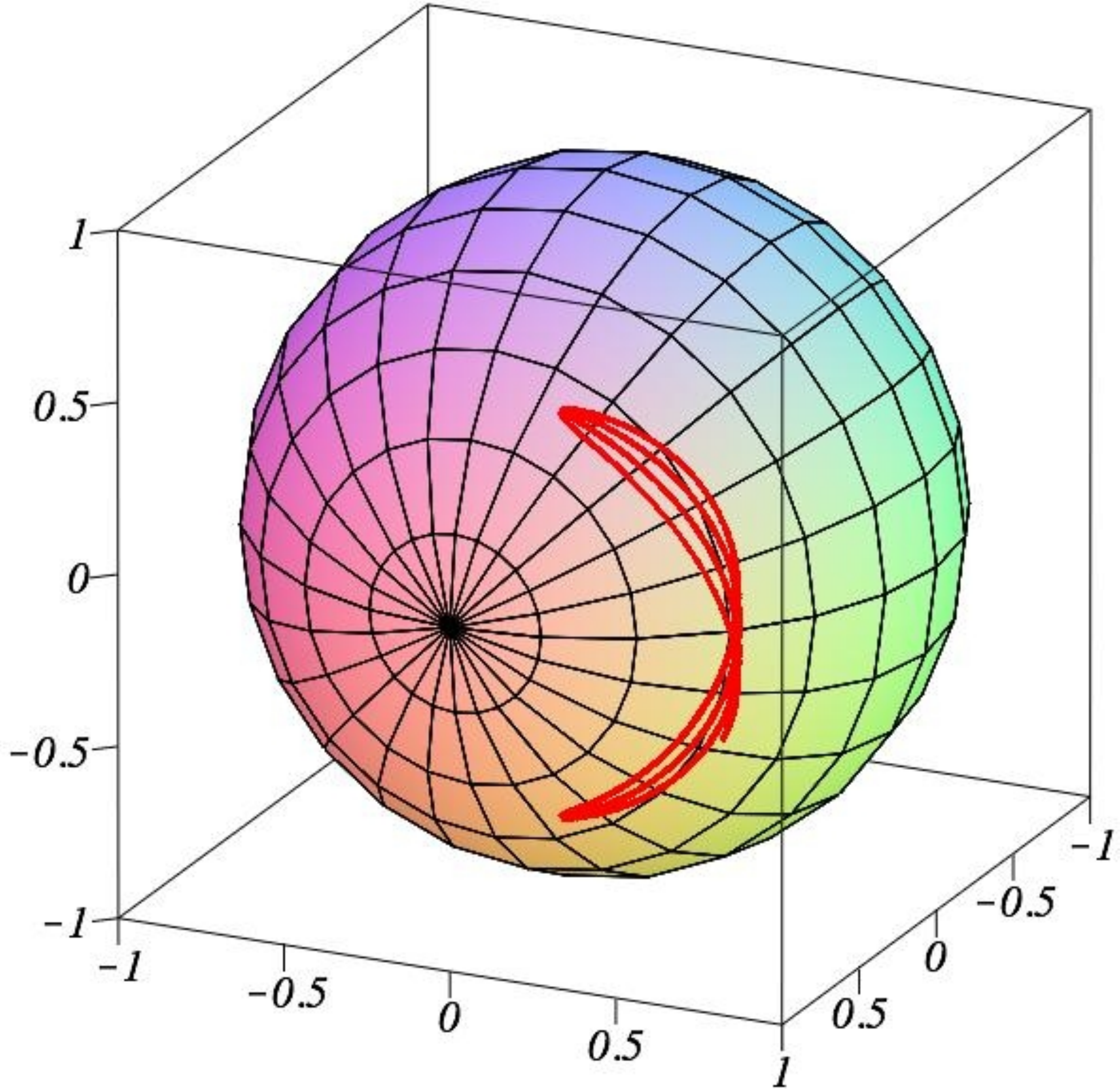}}
\put(103,72){\footnotesize $x_1$}
\put(132,76){\footnotesize  $x_0$}
\put(82,98){\footnotesize $x_2$}
\put(113,67){$(b)$}
\put(14,8){\includegraphics[scale=0.26]{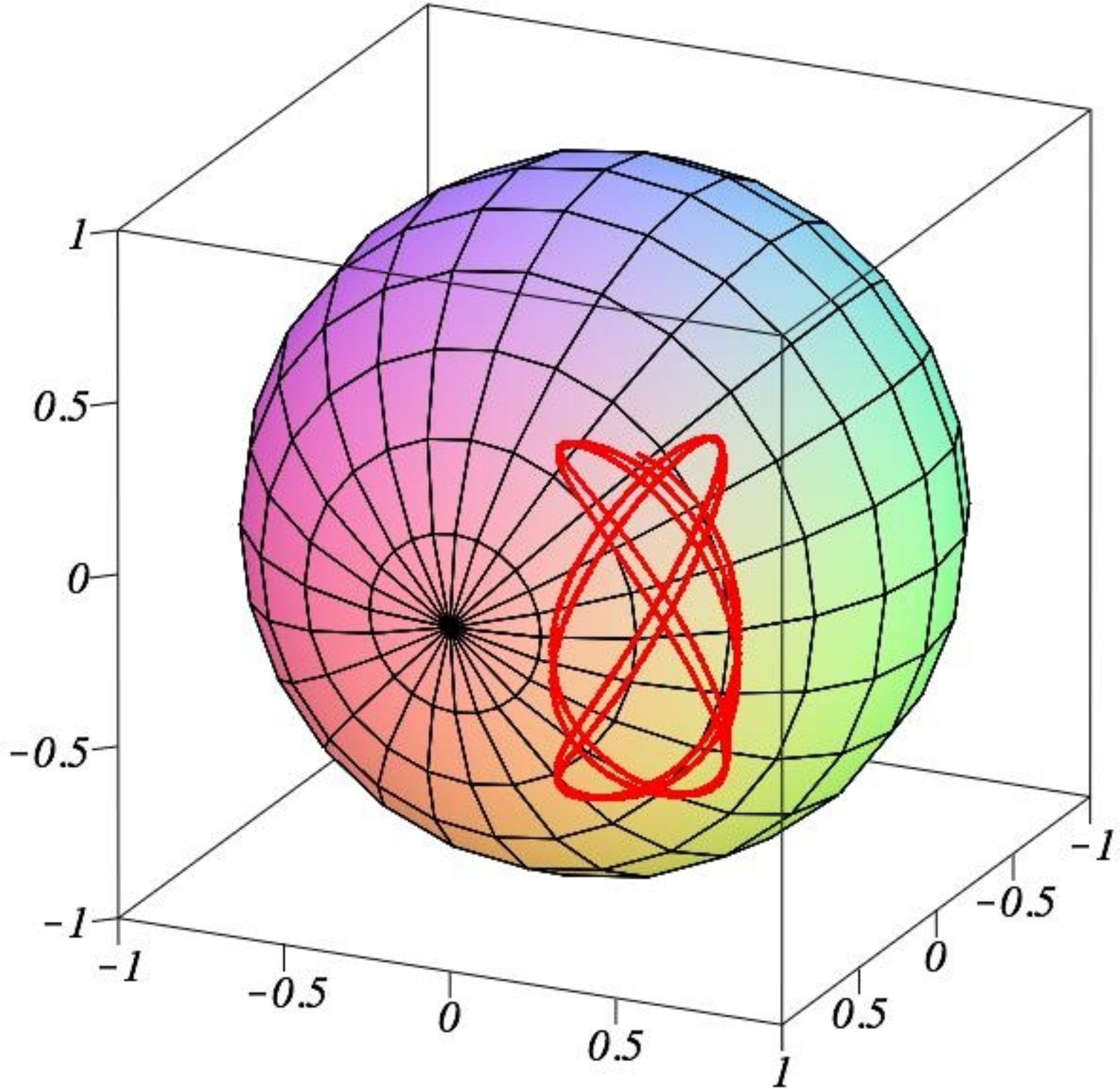}}
\put(33,7){\footnotesize $x_1$}
\put(62,11){\footnotesize  $x_0$}
\put(11,33){\footnotesize $x_2$}
\put(43,2){$(c)$}
\put(84,8){\includegraphics[scale=0.26]{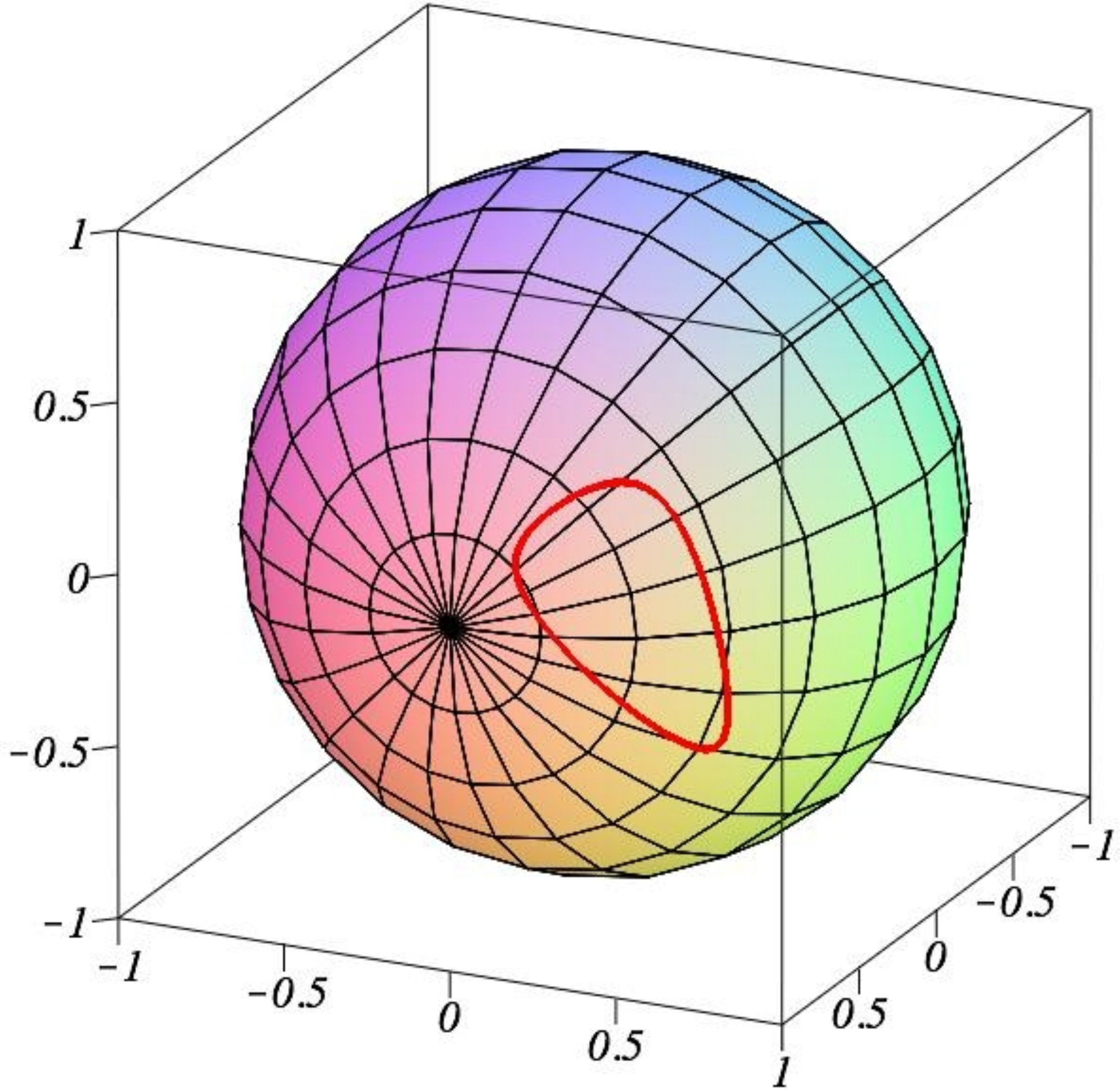}}
\put(103,7){\footnotesize $x_1$}
\put(132,11){\footnotesize $x_0$}
\put(81,33){\footnotesize $x_2$}
\put(113,2){$(d)$}
}
\end{picture}
\caption {\footnotesize Some trajectories for the Hamiltonians ${\cal H}_\kk$ (\ref{da}) and $\non_\kk$ (\ref{dd})  on    ${\mathbf S}^2$   ($\k=1$) with a single Rosochatius potential  such that $\lambda_1=0.1$ and $\lambda_2=0$. Time runs from $t=0$ to $t=8$ with the initial data $q_1=1$, $\dot{q}_1=1$, $q_2=-0.5$, 
$\dot{q}_2=2$:
$(a)$ the superintegrable case of ${\cal H}_\kk$ with $\delta=1$ and   $\Omega=0$ (Higgs oscillator with one centrifugal term),
$(b)$ the integrable ${\cal H}_\kk$ with $\delta=1$ and  $\Omega=0.2$,
$(c)$  the integrable  ${\cal H}_\kk$ with $\delta=1$ and  $\Omega=3$ and
$(d)$ the superintegrable   $\non_\kk$ with $\delta=1$.     }
\end{figure}
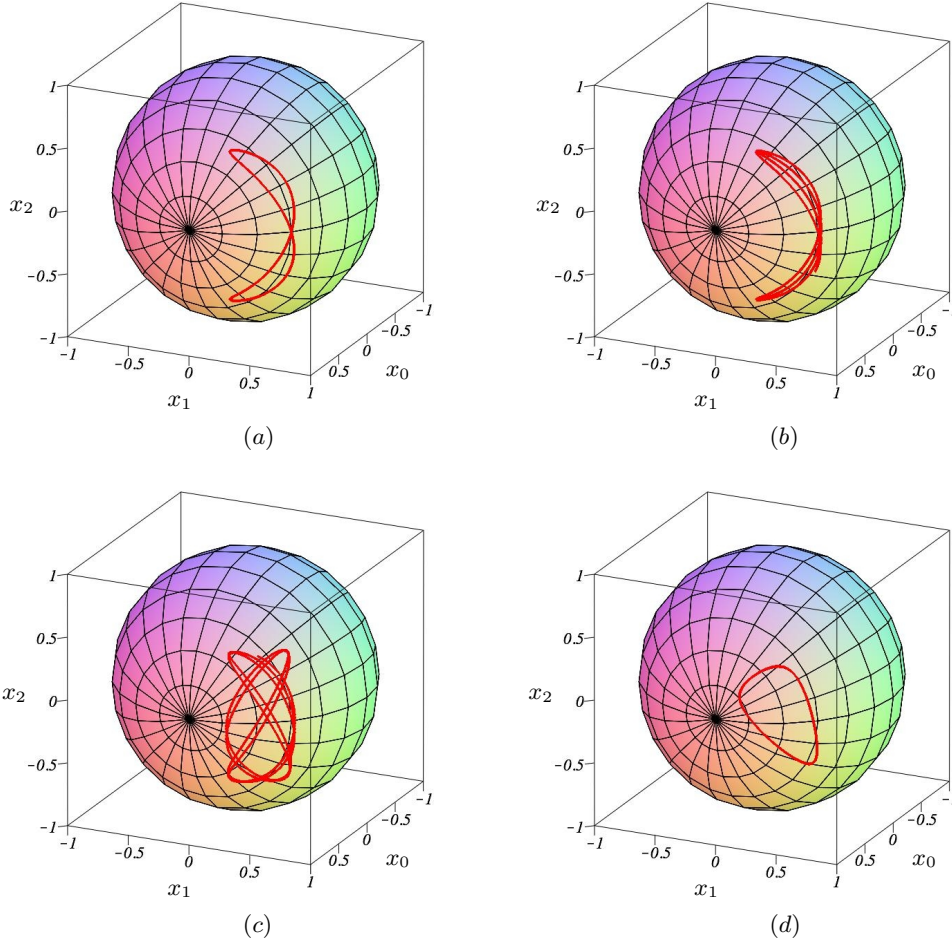


\subsect{Some trajectories}

In order to illustrate the above results and to highlight the role of the curved centrifugal terms,  we plot  in figure 6  some trajectories of the Hamiltonians ${\cal H}_\kk$ (\ref{da}) and $\non_\kk$ (\ref{dd})
on the sphere  ${\mathbf S}^2$  in ambient coordinates and with a single centrifugal term ($\la_1> 0$ and $\la_2=0$).  
These trajectories are numerically integrated by imposing the same initial conditions and values of the constants $\del$ and $\Om$ as in the trajectories plotted in figure 2 (which is the same system with no centrifugal term). Consequently, the effect of the $\la_1$-potential becomes apparent by comparing figures 2 and 6.
In this respect, it has to be taken into account that the $\la_1$-term is associated to the ambient coordinate $x_1$ (see (\ref{dddff})),  with center $O_1=(0,1,0)$ for the noncentral oscillator and geodesic radial distance $r_1$. In this way, we find that:
\begin{itemize}

\item The trajectories are restricted to the  `right'  semisphere due to the infinite barrier arising at $x_1=0$. Roughly speaking, this barrier induces a   `folding'  of the initial trajectories in figure 2.

\item  The superintegrable Higgs oscillator $(\Om= 0)$ with a centrifugal term provides a kind of Lissajous 1:2 curve (figure 6$(a)$).

\item  The integrable anisotropic systems    ${\cal H}_\kk$ with $\Om\ne 0$  represented in figures 6$(b)$ and 6$(c)$ show bounded trajectories which are again nonperiodic.

\item The  superintegrable $\non_\kk$ with a centrifugal term/noncentral oscillator gives rise to a  kind of Lissajous 1:1 curve (figure 6$(d)$).

\end{itemize}

Next, if the second centrifugal term with $\la_2> 0$ is also considered, then similar features arise from the second infinite barrier. For the sake of completeness, figure 7 contains the trajectories on the sphere of the Hamiltonian  ${\cal H}_\kk$ (\ref{da})   which exactly correspond to those plotted in figure 6 but now with the second   Rosochatius potential. We remark that:

\begin{itemize}

\item The trajectories of   ${\cal H}_\kk$ are   restricted to    a quadrant of the sphere;  these can be seen as  once  `folded'  with respect to the trajectories shown in figure 6 but twice  with respect to those obtained in figure 2.

\item  The superintegrable Higgs oscillator $(\Om= 0)$ with two centrifugal terms (curved SW system) gives rise to a Lissajous 1:1 curve (figure 7$(a)$).

\item  As expected once more,  the  trajectories coming from the integrable anisotropic Hamiltonian    ${\cal H}_\kk$ with $\Om\ne 0$  drawn in figures 7$(b)$ and 7$(c)$ show bounded nonperiodic trajectories.

\end{itemize}

Summarizing, we can say that the addition of the appropriate curved centrifugal terms preserves the essential dynamical features of all these systems.


\begin{figure}[t]
\setlength{\unitlength}{1mm}
\begin{picture}(140,120)(0,0)
\footnotesize{
\put(14,73){\includegraphics[scale=0.26]{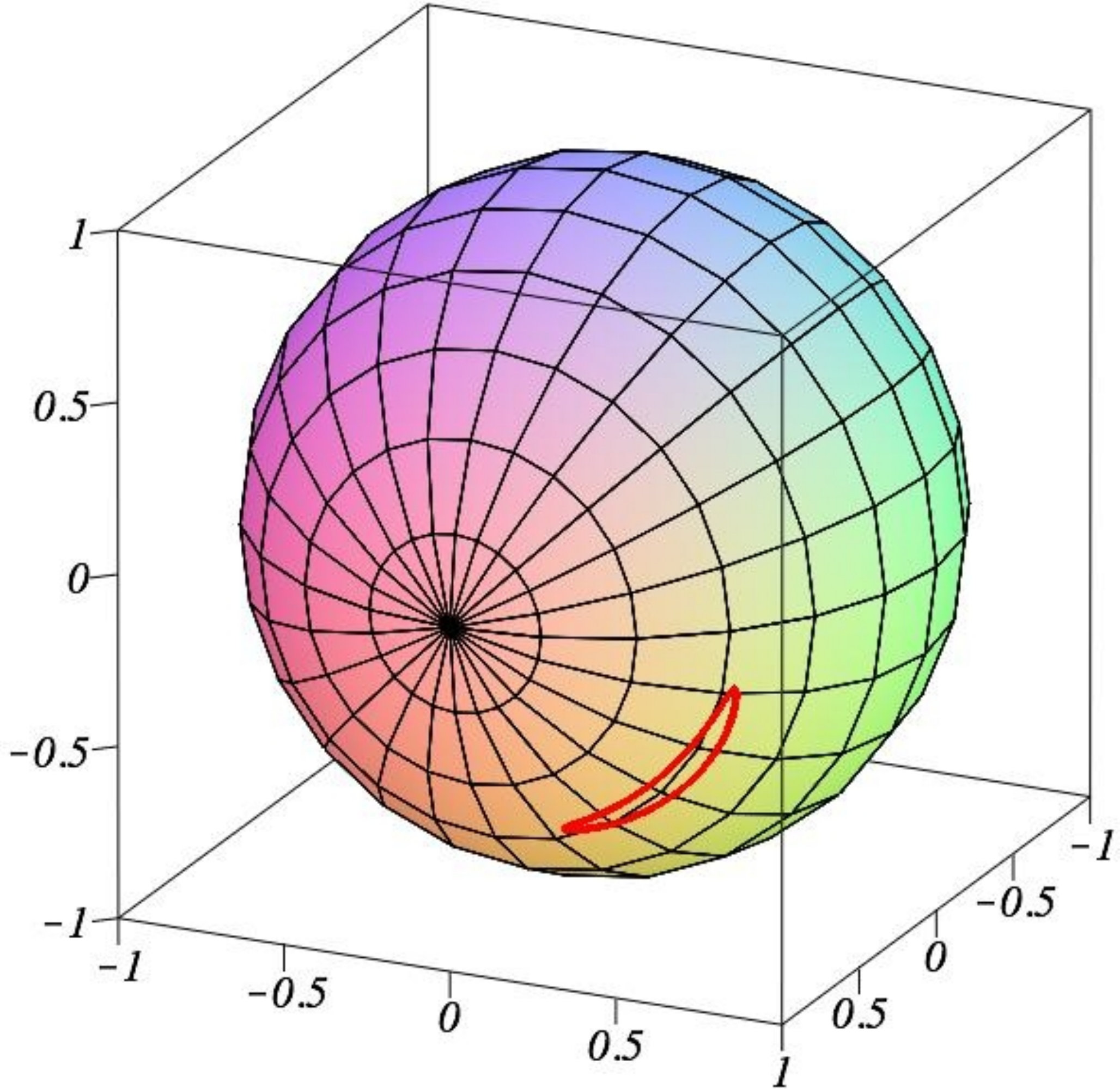}}
\put(33,72){\footnotesize{$x_1$}}
\put(62,76){\footnotesize  $x_0$}
\put(12,98){\footnotesize  $x_2$}
\put(43,67){$(a)$}
\put(84,73){\includegraphics[scale=0.26]{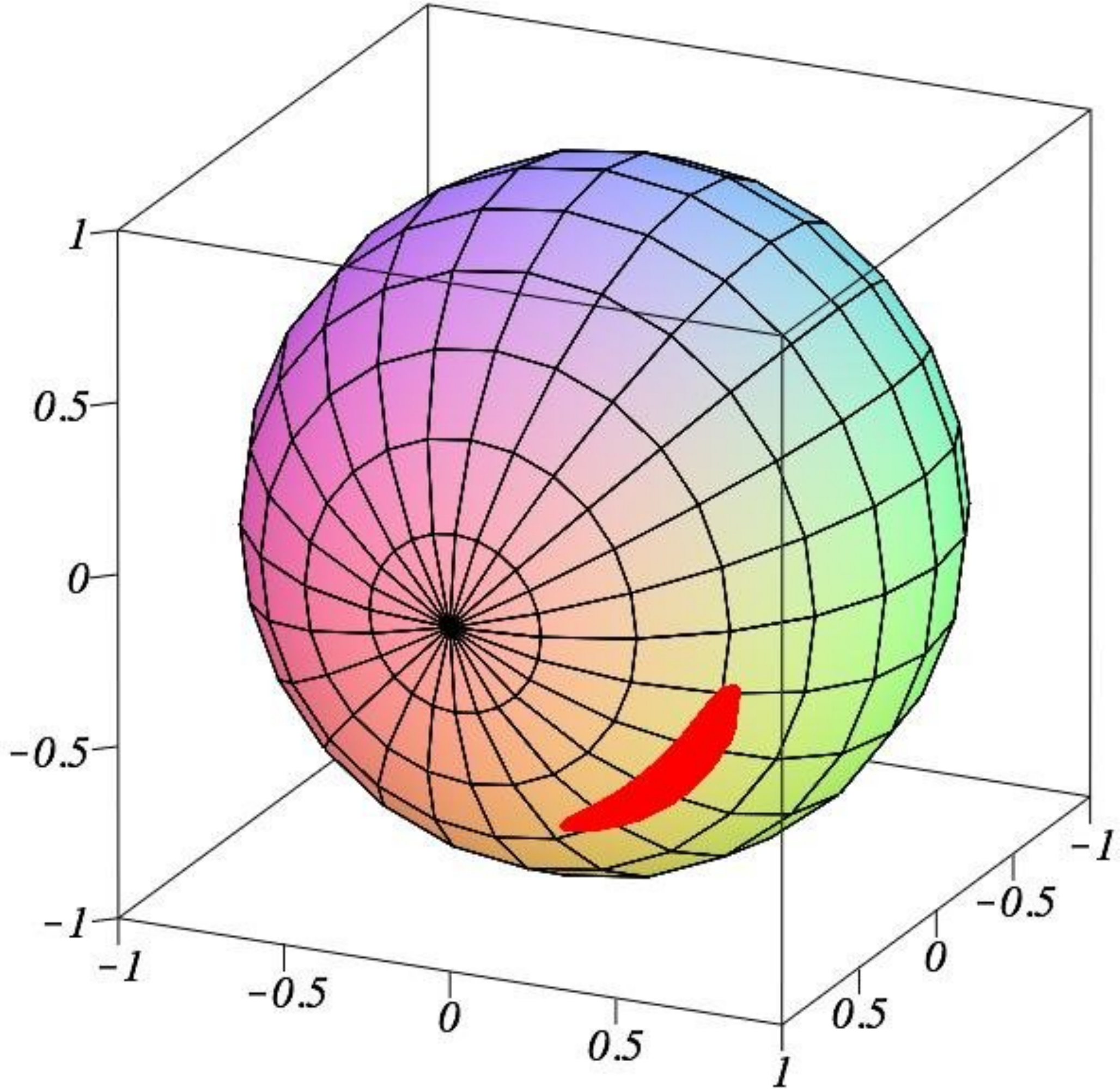}}
\put(103,72){\footnotesize $x_1$}
\put(132,76){\footnotesize  $x_0$}
\put(82,98){\footnotesize $x_2$}
\put(113,67){$(b)$}
\put(14,8){\includegraphics[scale=0.26]{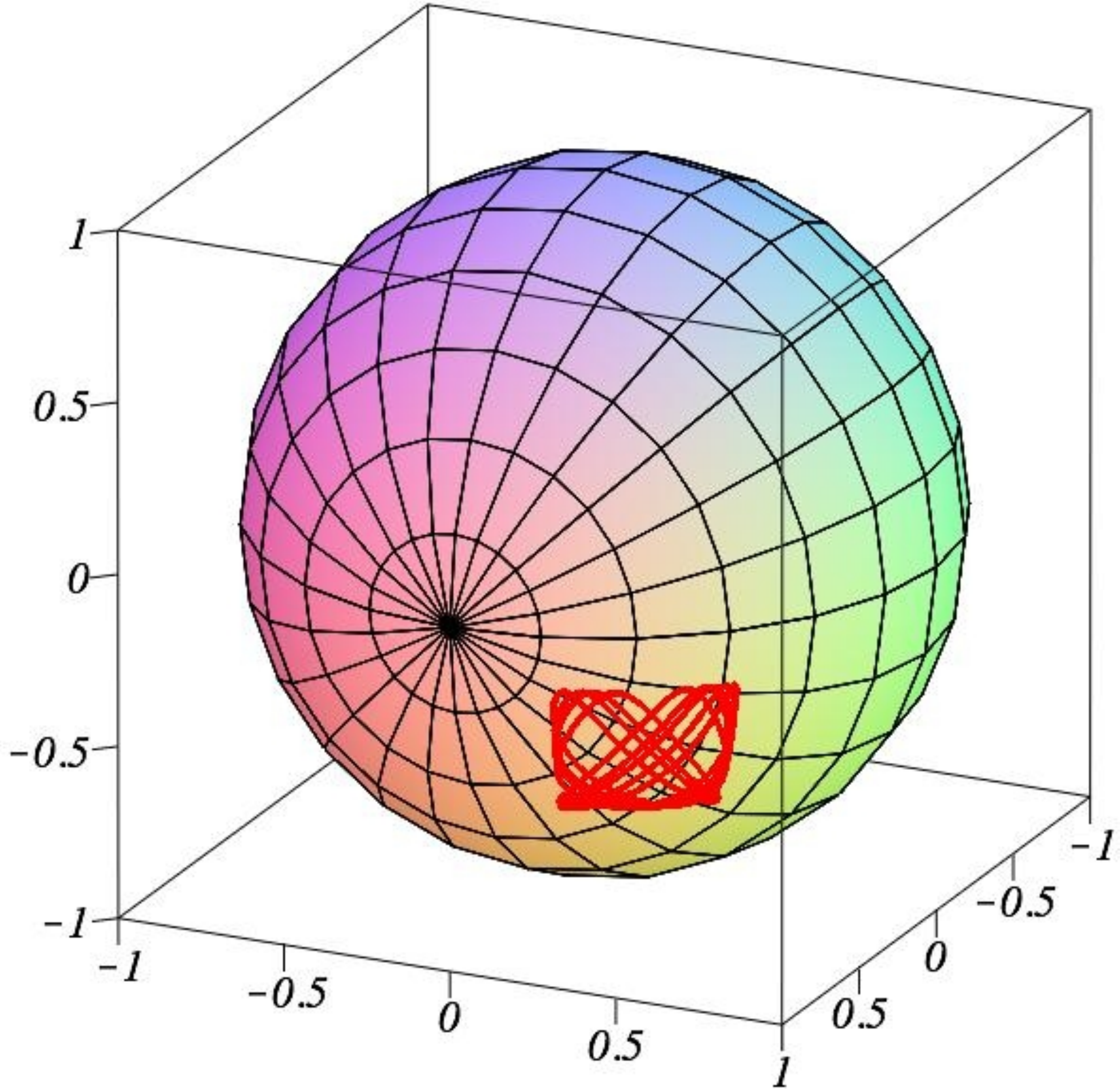}}
\put(33,7){\footnotesize $x_1$}
\put(62,11){\footnotesize  $x_0$}
\put(11,33){\footnotesize $x_2$}
\put(43,2){$(c)$}
}
\end{picture}
\caption {\footnotesize Trajectories of the Hamiltonian  ${\cal H}_\kk$ (\ref{da})  on ${\mathbf S}^2$    ($\k=1$)   with   two Rosochatius potentials given by $\lambda_1=0.1$ and $\lambda_2=0.05$. Initial data are the same as in figures 2 and 6:
$(a)$ the superintegrable case of ${\cal H}_\kk$ with $\delta=1$ and   $\Omega=0$ (Higgs oscillator with two centrifugal terms or curved SW system),
$(b)$ the integrable ${\cal H}_\kk$ with $\delta=1$ and  $\Omega=0.2$  and
$(c)$  the integrable  ${\cal H}_\kk$ with $\delta=1$ and  $\Omega=3$.  Note that  $\non_\kk$  is not defined when both $\lambda_1$ and $\la_2$ are non-zero. }
\end{figure}
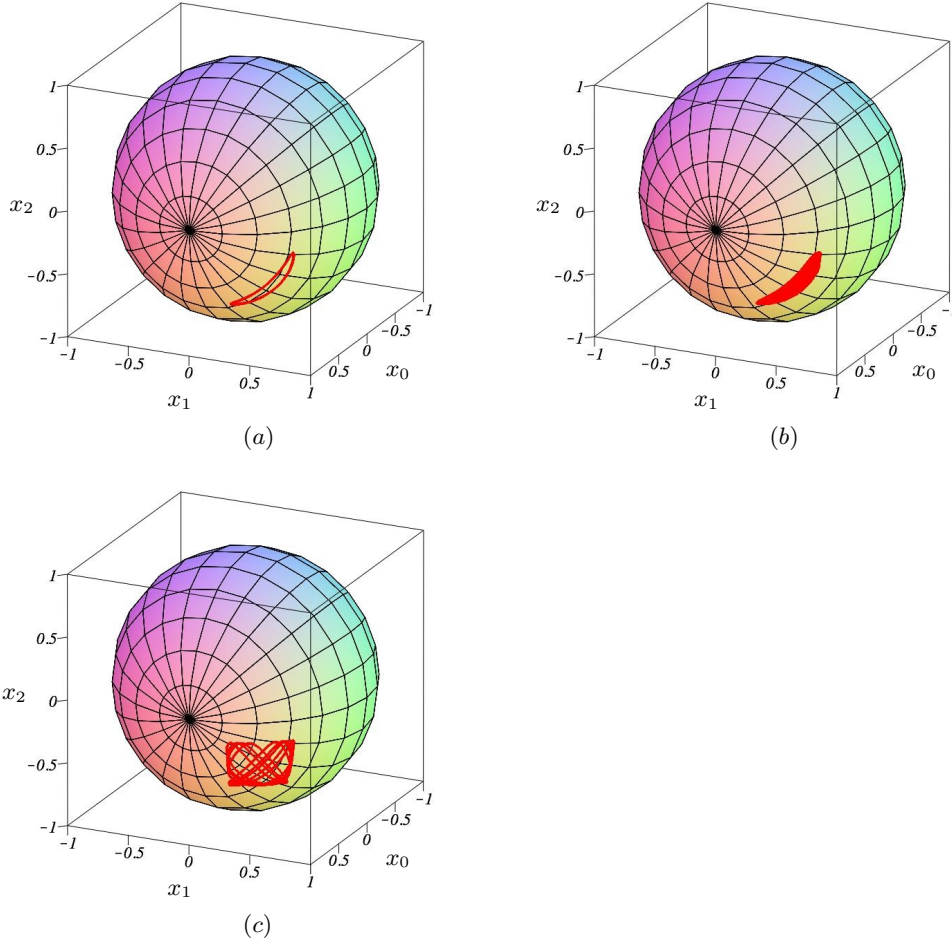



\section{Concluding remarks}

In this work we have introduced a new Hamiltonian  ${\cal H}_\kk$ on  ${\bf S}^2$ and  ${\bf H}^2$,  which is integrable for any value of the anisotropy parameter $\Om$ and that can be  interpreted as the  anisotropic Higgs oscillator      with two 
curved Rosochatius potentials. Furthermore, such a system together with the previously known curved superintegrable 1:2 oscillator $\non_\kk$ have been   algebraically and geometrically studied.  Numerical solutions for  the trajectories corresponding to some particular initial conditions for  both Hamiltonians have been  presented. In this respect, some remarks are in order.

\begin{itemize}

\item The flat contraction $\kk\to  0$ is always  a well defined and smooth limit in all the   expressions presented in  the paper. In this way the known systems presented in section 1 can always be  recovered in the Euclidean limit. Hence 
the   two different integrable Hamiltonians ${\cal H}_\kk$ and $\non_\kk$    can be thought as generalizations of  the known  Euclidean systems     (\ref{ba}) and  (\ref{bd}) (or (\ref{bf}) and (\ref{bg})), and $\non_{\kk=0}$  is nothing but a particular (superintegrable) case of ${\cal H}_{\kk=0}$.

\item Alternatively,  this approach can be  seen  as a  `curvature-deformation' process of  the Euclidean anharmonic oscillator
that by starting from   (\ref{bf}) and (\ref{bg}) arrives at the Hamiltonians (\ref{da}) and (\ref{dd}), respectively. 
In this sense,  we remark that there is no an apparent  straightforward prescription in order to perform such a deformation, as the explicit expressions for the curved Hamiltonians show.

\item 
It is also worth stressing that when $\kk\ne 0$ the superintegrable Hamiltonian (\ref{dd})   {\em cannot}  be recovered from the integrable one (\ref{da}) by setting $\Om=3\del$ and $\la_2=0$ (see also the   potentials (\ref{dddff}) as well as  the integrals ${\cal I}_{i,\kk}$ (\ref{db}) and  $\ii_{i,\k}$ (\ref{de})). This, in turn, means that  the Hamiltonians ${\cal H}_\kk$ and $\non_\kk$  are two {\em completely different} systems  on the sphere and on the hyperbolic plane. This fact indicates that different integrable curved systems may collapse under contraction into a single flat one and, conversely, that the integrable generalization of a known commensurate Euclidean Hamiltonian could split into several generalized   `curved'  possibilites.

\end{itemize}

Starting from all these results, several further research directions can be explored. Firstly, the search for superintegrable curved generalizations of other commensurate Euclidean oscillators should be faced. In this respect, we recall that the superintegrable Hamiltonian $\non$ (\ref{bd}) is the only system in the series of anisotropic oscillators with commensurate frequencies that is endowed with
  quadratic integrals of motion (see~\cite{Tempesta}). Moreover, the  `additional'  integral  that ensures the superintegrability of the system is, in fact, a second order function arising from a true sixth order integral. More explicitly, if we set $\la_1=0$  then it is known~\cite{Tempesta} that
$$
  Q_1=\left(p_1^2- 2 \del q_1^2 - 2{\rm i} \sqrt{2\del} q_1p_1  \right)^2\left(p_2^2- 8 \del q_2^2 +4{\rm i} \sqrt{2\del} q_2p_2  \right),\qquad \bar{Q}_1 ,\qquad Q=\frac 12 \left( Q_1+ \bar{Q}_1 \right),
$$
 are sixth order integrals of motion of $\non$, and that the set $ \ii_1$,  $\ii_2 $ (\ref{bd})  and $Q$ is formed by three functionally independent functions. Alternatively, the sixth order integral $Q$ can be written as
 $$
 Q=\left( p_1^2 +2\del q_1^2\right)^2\left( p_2^2 + 8 \del q_2^2\right)- 16\del\left( 2\del q_1^2q_2+(q_1 p_2 - q_2 p_1)p_1 \right)^2.
  $$
 Now, the second order integral $\ele$ (\ref{bd})  arises through the following relation
 $$
 {\ele}^2=\frac {8{ \ii_1}^2\ii_2-Q}{16\del}.
 $$
Remarkably enough, for the curved superintegrable Hamiltonian  $\non_\kk$ (\ref{de}) (in Beltrami variables), the corresponding  sixth order integral $Q_\k$ can also be  found, namely,
$$
Q_\k=\left(     J_{01}^2+\kk J_{12}^2      +2    \del\, \frac{q_1^2(1+\kk q_2^2) }{(1-\kk q_2^2)^2}     \right)^2\left( J_{02}^2 +     \frac{8 \del q_2^2  }{(1-\kk q_2^2)^2}   \right) - 16\del\left(  \frac{2\del q_1^2 q_2 }{(1-\kk q_2^2)^2}  +J_{12}J_{01}\right)^2,
$$
where $J_{ij}$ are the functions  (\ref{cd}). Moreover, the relationship with the second order integral $\ele_\k$ (\ref{de}) remains in the same  `flat' form  
  $$
 {\ele}_{\!\!\k}^2=\frac {8{ {\ii}^2_{1,\k}}\,\ii_{2,\k}-Q_\k}{16\del}.
 $$
Therefore these latter expressions seem to point out that 
the construction of the ${\bf S}^2$ and  ${\bf H}^2$ analogues of some other Euclidean commensurate oscillators could be feasible.

Secondly, another interesting open problem would be the construction of the 
 constant curvature analogues of some integrable H\'enon--Heiles systems (see~\cite{alfonso} and references therein)  that can be written as particular cases of the multiparametric family 
$$
\mathcal{H}=\frac{1}{2}(p_1^2+p_2^2)+ \del q_{1}^{2}+(\del + \Om)q_{2}^{2}+\alpha\left(q_1^2 q_2 +\beta q_2^3\right) +\frac{\la_1}{q_{1}^{2}} ,
$$
in which the only known integrable cases are the
 Sawada--Kotera   ($\beta=1/3, \Om=0$), the KdV   ($\beta=2, \Om$ arbitrary) and  the Kaup--Kupershmidt 
  ($\beta=16/3, \Om=15\del$) systems.
The results here presented would be a very reasonable starting point, that should be completed with appropriate curved analogues of the cubic H\'enon--Heiles term.

Finally, the explicit solution of the Schr\"odinger equation associated to ${\cal H}_\kk$ and $\non_\kk$ should be addressed, as well as the analysis of the quantum nonlinear dynamics generated by these new class of integrable nonlinear quantum models by following, e.g., the approaches presented in~\cite{ann1,ann2,ann3}. Work on all these lines is in progress.


\section*{Acknowledgments}

This work was partially supported by the Spanish MINECO under grants  MTM2010-18556 and AIC-D-2011-0711 (MINECO-INFN).



\begin{thebibliography}{10}\frenchspacing

\small



 



\bibitem{FMSUW65}
Fris J, Mandrosov V,  Smorodinsky Y A,  Uhlir M and  Winternitz P 1965  On higher symmetries in quantum mechanics
{\em Phys. Lett.} {\bf 16}  354--6


 

\bibitem{Ev90b}
 Evans N W 1990  Superintegrability of the  Winternitz system
{\em Phys. Lett. A} {\bf 147}  483--6


 

 \bibitem{8}
 Evans N W  1991  Group theory of the Smorodinsky--Winternitz system
{\em J. Math. Phys.}  {\bf 32}    3369--75





 \bibitem{10}
 Grosche C,  Pogosyan G S  and   Sissakian A N 1995 Path integral discussion for Smorodinsky--Winternitz potentials 
1. Two- and three-dimensional  Euclidean space
 {\em Fortschr. Phys.} {\bf 43}  453--521 
 
 
 \bibitem{Jauch}  Jauch  J M  and  Hill E L 1940  On the problem of degeneracy in quantum mechanics {\em Phys. Rev.} {\bf  57} 641--5
  
  
 
  \bibitem{Tempesta}
 Rodr\1guez M A,      Tempesta P and Winternitz P 2008 Reduction of superintegrable systems: The anisotropic harmonic oscillator  {\em Phys. Rev. E} {\bf 78} 046608  (6pp)
 

  
  
\bibitem{Higgs}
 Higgs P W  1979  Dynamical symmetries in a spherical geometry I
{\em J. Phys.  A: Math. Gen.}  {\bf 12}   309--23 



\bibitem{Leemon}
 Leemon   H I 1979
 Dynamical symmetries in a spherical geometry II
{\em J. Phys.  A: Math. Gen.}  {\bf 12}  489--501






\bibitem{Pogosyan1}
 Grosche C,  Pogosyan G S and   Sissakian A N 1995
  Path integral discussion for Smorodinsky--Winternitz  potentials
2. The two- and three-dimensional   sphere
{\em Fortschr. Phys.} {\bf 43}  523--63


 \bibitem{Pogosyan2}
 Kalnins E G,  Miller W Jr and    Pogosyan G S 1997
 Superintegrability   of the two-dimensional  
 hyperboloid
{\em J. Math. Phys.} {\bf 38} 5416--33




   \bibitem{RS}
 Ra\~nada M F and    Santander M 1999 Superintegrable systems on the two-dimensional sphere $S^2$ and
the hyperbolic plane $H^2$
{\em J. Math. Phys.} {\bf 40}    5026--57






\bibitem{int}
Ballesteros A, Herranz F J,    Santander M  and Sanz-Gil T 2003 Maximal superintegrability on N-dimensional curved spaces
{\em J. Phys. A: Math. Gen.} {\bf 36}   L93--9




\bibitem{kiev} Ballesteros A and  Herranz F J  2006 Superintegrability on three-dimensional Riemannian
and relativistic spaces of constant curvature
   {\em SIGMA} {\bf 2}    010 (22pp)

  


  



\bibitem{ann0}
 Ballesteros A,  Enciso A,   Herranz F J and   Ragnisco O 2009    Superintegrability on N-dimensional curved spaces:
Central potentials, centrifugal terms and monopoles  {\em Ann. Phys.}
{\bf 324}  1219--33
 





\bibitem{conf}
 Herranz F J and    Santander M  2002 Conformal symmetries of spacetimes
{\em J. Phys. A: Math. Gen.} {\bf 35}   6601--18

 

  


\bibitem{trigo}
 Herranz F J,  Ortega R and  Santander M  2000 Trigonometry of spacetimes: A new self-dual approach to a
      curvature/signature (in)dependent trigonometry
 {\em J. Phys.  A: Math. Gen.} {\bf 33}   4525--51 



  
\bibitem{Kepler}
Ballesteros A and Herranz F J  2009 Maximal superintegrability of the generalized
Kepler--Coulomb system on N-dimensional
curved spaces
{\em J. Phys. A: Math. Theor.}
{\bf  42} 245203 (12pp)







\bibitem{Ranran}
Cari\~nena J F,    Ra\~nada  M F,    Santander M  and  Senthilvelan  M 2004 A non-linear oscillator with quasi-harmonic
behaviour: two- and n-dimensional oscillators  {\em Nonlinearity}  {\bf 17}  1941-63 


\bibitem{Voz}
Vozmischeva T G 2003  Integrable problems of Celestial Mechanics in spaces of constant curvature, Astrophysics and Space Science Library, vol 295 (Dordrecht: Kluwer Academic Publishers)



   


\bibitem{kalnins}
Kalnins E G, Benenti S and Miller W Jr 1997 Integrability, St\"ackel spaces, and rational potentials {\em J. Math. Phys.} {\bf  38} 2345--65.
  
    \bibitem{Saksida}  Saksida P  2001  Integrable anharmonic oscillators on spheres and hyperbolic spaces   {\em    Nonlinearity} {\bf  14}  977--94



    
  
  \bibitem{ran}
    Ra\~nada  M F and    Santander M  2002  On some properties of harmonic oscillator on spaces
of constant curvature {\em Rep. Math. Phys.}  {\bf 49}   335--43  


 \bibitem{ran1}
 Ra\~nada M F and  Santander M 2002 On harmonic oscillators on the two-dimensional sphere $S^2$ and
the hyperbolic plane $H^2$ {\em J. Math. Phys.} {\bf 43}  431--51 


 \bibitem{ran2}
    Ra\~nada  M F and    Santander M  2003 On harmonic oscillators on the two-dimensional sphere $S^2$ and the hyperbolic plane $H^2$ II
  {\em J. Math. Phys.}  {\bf 44}  2149-67 


\bibitem{CRMVulpi}
   Herranz F J,  Ballesteros A,  Santander M and   Sanz-Gil T 2004
Maximally  superintegrable   
Smorodinsky--Winternitz systems on the $N$-dimensional sphere and
hyperbolic spaces  {\em Superintegrability in Classical and Quantum  
Systems} 
Ed   P Tempesta {\em et al} 
{CRM Proceedings and Lecture Notes} vol 37 (Providence RI:   AMS)
pp 75--89 
 
 
 

 

\bibitem{alfonso}
Ballesteros A and Blasco A 2010 Integrable H\'enon--Heiles Hamiltonians:
A Poisson algebra approach
{\em Ann. Phys.}
{\bf  325} 2787--99

 


 \bibitem{ann1}
 Cari\~nena J F,  Ra\~nada M F and Santander M 2007
A quantum exactly solvable non-linear
oscillator with quasi-harmonic behaviour
{\em Ann. Phys.}
{\bf 322}  434--59

  \bibitem{ann2}
 Cari\~nena J F,  Ra\~nada M F and Santander M  2007
The quantum harmonic oscillator on the sphere
and the hyperbolic plane,
{\em Ann. Phys.}
{\bf 322}  2249--78


 

\bibitem{ann3}
 Ballesteros A, Enciso A,  Herranz F J,  Ragnisco O and    Riglioni D 2011  
Quantum mechanics on spaces of nonconstant curvature:
The oscillator problem and superintegrability
{\em Ann. Phys.}
{\bf 326}  2053--73
 


 

\end{thebibliography}
\end{document}